\documentclass[11pt,a4paper]{article}
\pdfoutput=1
\usepackage{jheppub}
\usepackage{latexsym,amsfonts,amsmath,amssymb,mathrsfs}
\usepackage{bbm}
\usepackage{graphicx}
\usepackage{amsmath}
\usepackage{mathtools}
\usepackage{color}
\usepackage{caption}
\usepackage{subcaption}
\usepackage[normalem]{ulem}
\usepackage{comment}
\usepackage{url}
\usepackage{slashed}
\usepackage{tabu}
\usepackage{multirow}
\usepackage{float}
\usepackage{tikz}
\usetikzlibrary{shapes}
\usepackage{makecell}

\renewcommand{\arraystretch}{2}       
\renewcommand{\arraystretch}{1.5}   
\newcommand{\be}{\begin{equation}}
\newcommand{\ee}{\end{equation}}
\newcommand{\ba}{\begin{array}}
\newcommand{\ea}{\end{array}}
\newcommand{\beqa}{\begin{eqnarray} }
\newcommand{\eeqa}{\end{eqnarray} }
\newcommand{\nn}{\nonumber}

\newcommand\cA{{\cal A}}

\newcommand\cC{{\cal C}}
\newcommand\cD{{\cal D}}

\newcommand\cF{{\cal F}}
\newcommand\cG{{\cal G}}

\newcommand\cI{{\cal I}}

\newcommand\cL{{\cal L}}
\newcommand\cM{{\cal M}}
\newcommand\cN{{\cal N}}
\newcommand\CN{{\cal N}}

\newcommand\cQ{{\cal Q}}

\newcommand\cV{{\cal V}}
\newcommand\cY{{\cal Y}}

\newcommand\gmr{{\phi(2\rho)}}

\newcommand{\R}{\ell}

\newcommand\btau{{\boldsymbol\tau}}
\newcommand\bsigma{{\boldsymbol\sigma}}
\newcommand\boa{{\boldsymbol{a}}}
\newcommand\bob{{\boldsymbol{b}}}
\newcommand\boc{{\boldsymbol{c}}}
\newcommand\ta{{\text{a}}}
\newcommand\tb{{\text{b}}}

\newcommand\tm{{\text{m}}}

\newcommand{\defeq}{\mathrel{\mathop:}=}

\newcommand\half{{\textstyle{\frac{1}{2}}}}

\def\i{\mathrm{i}}
\def\ve{\varepsilon}

\def\wt{\widetilde}

\renewcommand{\=}{\;  = \;}

\newcommand\ndt{{\noindent}}

\newcommand{\IZ}{\mathbb{Z}}

\newcommand{\IH}{\mathbb{H}}

\renewcommand{\t}{\tau}

\newcommand{\Tr}{{\rm Tr}}
\newcommand{\rme}{{\rm e}}

\newcommand{\adss}{\text{AdS}_3  \! \times \! \text{S}^2}
\newcommand{\Hthrees}{\IH^3/\IZ  \!\times \! \text{S}^2}

\title{Localization on $\adss$~I:
the 4d/5d connection in off-shell Euclidean supergravity}

\author{Axel Ciceri$^{a}$, Imtak Jeon$^{b,c}$, Sameer Murthy$^{a}$}

\affiliation{$^a$ Department of Mathematics, King's College London,\\
  The Strand, London WC2R 2LS, U.K}
\affiliation{$^b$ Asia Pacific Center for Theoretical Physics, Postech, Pohang 37673, Korea }
  \affiliation{$^c$ Department of Physics, Postech, Pohang 37673, Korea  }

\emailAdd{imtakjeon@gmail.com}
\emailAdd{sameer.murthy@kcl.ac.uk}
\emailAdd{axel.ciceri@kcl.ac.uk}

\abstract{
We begin to develop the formalism of localization for the functional integral 
of supergravity on~$\text{AdS}_3 \times  \text{S}^2$. 
We show how the condition of supersymmetry in the Euclidean~$\Hthrees$ geometry 
naturally leads to a twist of the~S$^2$ around the time direction of~AdS$_3$.
The twist gives us a five-dimensional Euclidean supergravity background dual to the elliptic 
genus of~$(0,4)$ SCFT$_2$ at the semiclassical level. 
On this background we set up the off-shell BPS equations for one of the Killing spinors, 
such that the functional integral of five-dimensional Euclidean supergravity on~$\Hthrees$ 
localizes to its space of solutions.
We obtain a class of solutions to these equations by lifting known off-shell BPS solutions of  
4-dimensional supergravity on~AdS$_2 \! \times \! \text{S}^2$. 
In order to do this consistently, we construct and use a Euclidean version of the off-shell 4d/5d lift of 
\href{https://arxiv.org/abs/1112.5371}{{\ttfamily arxiv:1112.5371}}, which could be of 
independent interest. 
}

\begin{document}

\voffset 0in

\maketitle

\voffset -1in



\section{Introduction}

In this paper we take the first steps to develop the formalism of localization in supergravity 
on asymptotic $\adss$.
Our main physical motivation is to study an exact sector of AdS$_3$/CFT$_2$ holography. 
By exact, we mean the inclusion of all quantum effects in supergravity or, equivalently, all finite-charge 
corrections to the results derived in the limit of infinite central charge of~CFT$_2$. 
Our eventual goal is to calculate observables protected by supersymmetry in the bulk gravitational theory using 
localization applied to functional integrals in supergravity. The same approach has yielded rich results in 
theories on~AdS$_2$ (times~S$^2$ or~S$^3$) culminating in the calculation of the exact quantum entropy 
of black holes in four dimensions~\cite{Dabholkar:2010uh,Dabholkar:2011ec,Dabholkar:2014ema,Iliesiu:2022kny}
and five-dimensions~\cite{Gupta:2021roy}.
The idea is to extend such calculations to higher dimensional AdS spaces. 

In this paper we focus on the AdS$_3$/CFT$_2$ context and consider
the bulk~AdS$_3$ calculation of the elliptic genus of the dual SCFT$_2$. 
Many of the intermediate calculations are motivated by the embedding of this problem in string theory, wherein 
two-dimensional SCFTs are realized as the low-energy theory on the world-volume of effective strings.
Some canonical examples are the D1/D5 system in IIB string theory
wrapped on~$T^4$/$K3$~\cite{Strominger:1996sh, Maldacena:1999bp}, 
$M5$-brane bound states in M-theory/CY$_3$~\cite{Maldacena:1997de},
and D3 branes wrapping curves on the base of elliptically fibered CY$_3$ in F-theory~\cite{Vafa:1997gr}.  
All these SCFTs have at least~$(0,4)$ supersymmetry in two dimensions with a corresponding~$SU(2)$ R-symmetry.
Focussing on the MSW theory~\cite{Maldacena:1997de}, the gravitational dual in the generic theory has the 
form~$\adss \times$CY$_3$.\footnote{One has 
AdS$_3 \times$S$^3 \times T^4/K3$ in the D1/D5 system and 
AdS$_3 \times$S$^3 \times$CY$_3$ in the F-theory case~\cite{Haghighat:2015ega}.
In either of these cases, we consider the~$SU(2)\times \widetilde{SU(2)}$ action on~S$^3$, 
and we think of the~S$^2$ with~$\widetilde{SU(2)}$ action as being embedded in the~S$^3$.} 
\footnote{One could also consider black strings in AdS$_5$ which needs a different treatment using 
gauged supergravity which we will not consider here.} 

The low energy theory is summarized by five-dimensional supergravity with asymptotic $\adss$  
boundary conditions, which is the theory that we study in this paper in the off-shell conformal formalism.
The classical Lorentzian~$\adss$ theory in this formalism has been studied in a series of nice 
papers~\cite{Castro:2007sd,Castro:2007hc,Castro:2008ne}. 
Our eventual goal is to calculate functional integrals for quantum observables using localization 
in the corresponding Euclidean supergravity theory defined on global~$\adss$
with a periodic Euclidean time coordinate, i.e., the manifold~$\Hthrees$. 
The three-dimensional part of the manifold has the topology of a solid torus, and we often refer to it as such.

The localization problem in supergravity is substantially more difficult compared to its counterpart in 
quantum field theories. To begin with, one defines a rigid supercharge in the gravitational theory, 
using the background field method in theories with soft gauge algebras as applied to 
supergravity~\cite{deWit:2018dix,Jeon:2018kec}. 
At a practical level the problem reduces to finding all bosonic gravitational configurations that admit a Killing spinor
whose asymptotic limit is one of the supercharges of~$\adss$. 
Secondly, one finds all matter configurations invariant under this supercharge. 
Thirdly, one evaluates the supergravity action at a generic point in this manifold. Finally one calculates 
the one-loop determinant of the non-BPS fluctuations around the localization manifold. 

\medskip

In this paper we address the first two questions whose solutions comprise the so-called
localization locus. We work in the context of 5d off-shell~$\CN=1$ supergravity 
coupled to an arbitrary number of vector multiplets. 
Our idea is to use the 4d/5d lift~\cite{Banerjee:2011ts}, which relates solutions of off-shell 4d 
supergravity to those of off-shell 5d supergravity compactified on a circle.\footnote{This is different 
from the 4d/5d lift of~\cite{Gaiotto:2005gf} which involves a lift on a Taub-NUT space.} 
The localization manifold in 4d~$\CN=2$ supergravity on asymptotically~AdS$_2 \times$~S$^2$
has been completely determined, and we can lift those solutions 
to~$\adss$ . Although this is not guaranteed to produce all supersymmetric solutions, 
it should give all solutions that are independent of the circle of compactification.
Similar ideas have been 
used successfully to make progress in localization on~AdS$_2 \times$~S$^3$ theories 
in~\cite{Gomes:2013cca}, \cite{Gupta:2019xac,Gupta:2021roy}.

As it turns out, implementing this idea is not quite straightforward. 
Firstly, the 4d/5d map in~\cite{Banerjee:2011ts} is given for Lorentzian backgrounds while we 
need it for Euclidean backgrounds. To this end, we first work out a consistent set of
supersymmetry transformations in the five-dimensional Euclidean supergravity theory. 
In four dimensions we use the Euclidean supergravity discussed 
in~\cite{Cortes:2003zd,Cortes:2005uq,Cortes:2009cs,deWit:2017cle}, \cite{Jeon:2018kec}. 
We then map off-shell BPS solutions of the 4d theory to off-shell BPS solutions of the 5d theory.
Here there is an additional problem, namely that the 4d Euclidean theory carries a redundancy 
of allowed reality conditions which has no counterpart in the 5d Euclidean theory.
We show that this redundancy can be absorbed in a parameter whose role is to 
implement the symmetry breaking~$SO(1,1)_R \to \mathbb{I}$.\footnote{This parameter
is the Euclidean analog of the parameter that enforces~$SO(2)_R \to \mathbb{I}$ in~\cite{Banerjee:2011ts}.}
The second problem has to do with the global identifications of the background that we are interested 
in, i.e.~$\Hthrees$, which is not a Kaluza-Klein lift of Euclidean AdS$_2 \times$~S$^2$. 
The Kaluza-Klein condition was used in~\cite{Banerjee:2011ts} for the off-shell 4d/5d lift and, indeed, 
general off-shell configurations do not consistently lift from Euclidean AdS$_2 \times$~S$^2$ to~$\Hthrees$. 
Nevertheless, the class of off-shell solutions relevant for the 4d black hole problem  
can be lifted to the supersymmetric~$\IH_3/\IZ$, due to their enhanced rotational symmetry.

In this manner we obtain an adaptation of the 4d/5d lift relevant for the Euclidean AdS$_3$/CFT$_2$ problem, whose 
details we work out in Section~\ref{section:Euc4d5dlift}. 
The final part of the paper, in Section~\ref{section:locresults}, is an application of this lift to find a class of 
off-shell solutions in the~AdS$_3$ theory which contribute to the elliptic genus problem in the boundary SCFT$_2$. 
In a sequel to this paper we study the subsequent steps in the localization problem, in particular the 
5d supergravity action and its associated boundary terms. 

There is another subtlety that appears in the Euclidean functional integral realization of the supersymmetric 
observables in the AdS$_3$/CFT$_2$ context even before beginning the 4d/5d lift.
Consider a 2d $(0,4)$ superconformal field theory SCFT$_2$ on the boundary torus with complex structure~$\tau$. 
The elliptic genus can be defined as a trace over the Hilbert space on~S$^1$ 
in the Ramond sector with an insertion of~$(-1)^F$,
which can be calculated by reducing to the free theory or by using index theorems~\cite{Witten:1986bf,Schellekens:1986xh,Alvarez:1987wg}. 
In the functional integral formalism, one correspondingly chooses
periodic boundary conditions for the fermions around both cycles of the torus. 
This leads to a constant Killing spinor on the torus, which is used to localize the SCFT$_2$
functional integral~\cite{Benini:2013nda}. 
Now consider the calculation of the same functional integral in the bulk theory. 
The bosonic vacuum configuration is that of thermal~AdS$_3$. 
One of the circles (the space circle in the thermal~AdS$_3$) is contractible and therefore 
the fermions should have half-integer momentum around that circle at infinity.
Demanding that a spinor obeys the Killing spinor equation forces its momenta around the two cycles to be equal,
so that it has fixed non-zero momentum also along the non-contractible (time) direction. 
On the other hand, the non-contractible direction has  
periodicity~$\text{Im}(\tau)$ which is arbitrary, and therefore the above spinor with non-zero momentum 
is not well-defined on the torus.
This problem can be resolved by turning on 
a twist of the~S$^2$ around the non-contractible circle in~AdS$_3$, which allows for spinors 
which are constant in time and therefore well-defined. 
As we discuss in Section~\ref{EuclBackground}, this allows us to set up a supersymmetric background of the 
form~$\IH^3/\IZ \times$~S$^2$.\footnote{A related supersymmetric set-up has been discussed in the literature 
in the context of supersymmetric black holes in~AdS space~\cite{Cabo-Bizet:2018ehj} and, in 
particular, for BTZ black holes in~\cite{Larsen:2021wnu}.} 
The bulk calculation is the NS-sector calculation of the elliptic genus, which is 
equivalent to the boundary Ramond sector trace by a spectral flow in the~$(0,4)$ algebra. 
We show how the twist reduces the final asymptotic algebra to be a sub-algebra of the 
Brown-Henneaux-Coussaert~\cite{Brown:1986nw,Coussaert:1993jp}~$(0,4)$ algebra on AdS$_3$.

\smallskip

We note that our study of localization of supergravity on~$\adss$  has similarities with 
localization of 3d supersymmetric gauge theories on~AdS$_3$, which has been studied in \cite{Assel:2016pgi}. 
Indeed the analysis of asymptotic boundary conditions on the gauge fields is very similar in both 
cases. The main difference and new challenge (apart from the two extra dimensions) is that the metric also fluctuates in 
our analysis, as can be seen in our localization solutions. 
We also note that the idea of exact holography has also been explored in the literature in contexts different to 
the one that we study here. This includes Koszul duality~\cite{Costello:2020jbh}, topological 
string theory~\cite{Brennan:2017rbf}, and classical (large-$N$) gravitational theories~\cite{BenettiGenolini:2019jdz}. 
It would be interesting to explore possible connections with these approaches.

\smallskip

The plan of this paper is as follows. 
In Section~\ref{section:setup} we discuss five-dimensional supergravity coupled to 
matter multiplets. 
We then describe the classical $\adss$ solution in Lorentzian signature, and 
the Killing spinors and superalgebra on this background. 
In Section~\ref{EuclBackground} we present the Euclidean $\Hthrees$ solution, the Killing
spinors and superalgebra on this background, and the relation of the path integral to the trace definition  
of the elliptic genus. 
In Section~\ref{section:Euc4d5dlift} we present the off-shell 4d/5d map modified to the Euclidean 
signature and present the lift of~AdS$_2 \times$S$^2$ to~$\Hthrees$.
In Section~\ref{section:locresults} we apply our formalism to lift localization solutions 
from~AdS$_2 \times$S$^2$ localization solutions on~$\Hthrees$. 
In various appendices, we present the spinor and Clifford algebra conventions, the Killing spinors, 
Killing vectors, and Lie brackets of global~$\adss$, and the four- and five-dimensional 
Euclidean supersymmetry transformations in the respective conformal supergravities.

\section{Background and set-up of the problem} \label{section:setup} 

In this section we discuss five-dimensional supergravity coupled to matter multiplets. 
We show that the Euclidean theory is obtained by redefinitions of fields of the Lorentzian theory
that follow simply from the Wick rotation. 
We review the elements of the supergravity theory including the supermultiplets and the supersymmetry
transformations. 
We then describe the classical $\adss$ solution in Lorentzian signature, and 
present the Killing spinors on this background and the consequent superalgebra.

\subsection{Off-shell 5d supergravity}\label{5dSUGRA}

Off-shell supergravity in the superconformal formalism in Lorentzian signature in various dimensions 
has been known for many decades (see the book~\cite{Freedman:2012zz}).
Euclidean supergravity, on the other hand, is a less-studied subject and few references 
exist (e.g.~\cite{Cortes:2003zd,Cortes:2005uq,Cortes:2009cs,deWit:2017cle}).
In these references the method of time-like reduction from a higher-dimensional Lorentzian theory 
is used to systematically construct the Euclidean-signature theory. 
In this section we discuss the formalism of 5d conformal supergravity with $\CN=1$
(minimal) supersymmetry, i.e., 8 supercharges. 
The Lorentzian theory was constructed in~\cite{Bergshoeff:2001hc, Bergshoeff:2002qk, 
Bergshoeff:2004kh}, and in~\cite{Kugo:2000hn, Kugo:2000af, Fujita:2001kv}, and it is reviewed 
in the more recent~\cite{deWit:2009de, Banerjee:2011ts} whose conventions we follow. 
One potential systematic approach to construct the Euclidean theory would be to perform a timelike 
reduction on a 6d theory. However, we use a less formal approach here: we start from a Wick rotation 
and make an appropriate set of transformations on all the fields of the Lorentzian theory 
so that we obtain a consistent 5d Euclidean theory.\footnote{Indeed, such an approach also works 
successfully in four dimensions and the result agrees with the timelike reduction~\cite{Jeon:2018kec}.}

The starting point is the usual Wick rotation~$t = -\i  t_E$ relating the Lorentzian and Euclidean time coordinates.
This is followed by the appropriate transformations of all tensors for this coordinate change. 
Accordingly, the time directional gamma matrices are related by~$\gamma_{{t}} = \i \gamma_{{t}_E}$. 
Changing the signature of spacetime by this Wick rotation, in general, demands changing the nature 
of irreducible spinors. For instance, in four dimensions, while the Majorana representation of irreducible 
spinors is allowed in the Lorentzian theory, it is not the case in the Euclidean theory. 
Therefore, we need an appropriate field redefinition of spinors. This can be achieved in a simple manner 
by going to the symplectic-Majorana 
basis which exists in both the Lorentzian and Euclidean theory. In this basis, the charge conjugation matrix 
is the same in both the theories. Therefore the Euclidean action follows from the Lorentzian action 
in this basis by simply implementing the Wick rotation
(See Appendix~B.4 of~\cite{Jeon:2018kec} for details of this map in four dimensions.)
Under the Wick rotation, the Lagrangian density is unchanged, 
i.e., $\cL_{\text{Lor.}} = \cL_{\text{Eulc.}}$, and so the action of the Lorentizan theory and Euclidean theory 
are related by~$\i S_{\text{Lor.}} = S_{\text{Eucl.}}$. 
Our conventions are such that the Euclidean action is to be negative definite when we impose well defined path integral.

The reality property of the fermionic fields in the Lorentzian and Euclidean theories are different. 
In the Lorentzian theory, an SU$(2)_R$ spinor doublet $\psi^i$ with $i= 1,2$ follows 
a symplectic-Majorana condition, as reviewed in~\eqref{SMLor},
\be
(\psi^i)^\dagger \gamma_{\hat{t}} \= \varepsilon_{ij}(\psi^j)^T C \,,
\ee
where~$C$ is the unique choice of the charge conjugation matrix in five dimensions 
(this is more generally true in odd dimensions). 
Compatibility with supersymmetry leads to the standard reality property on fluctuating bosons, 
in which the gauge fields and the metric are real. 
In the Euclidean theory the allowed reality condition for the spinors  
\be
(\psi^i)^\dagger \= \varepsilon_{ij}(\psi^j)^T C 
\ee
is not compatible with supersymmetry if we impose the usual reality conditions for fluctuating bosons. 
Therefore, we treat $\psi^1$ and $\psi^2$ as two independent Dirac spinors, formally doubling the fermionic 
degrees of freedom and then choosing a middle-dimensional contour in the functional integral, following 
the standard treatment of fermions in the Euclidean theory. This allows us to impose the standard reality conditions 
on bosonic fluctuations, so that the Euclidean action is negative-definite and the 
functional integral is well-defined at the perturbative level.\footnote{Note that we still allow reality conditions on the background 
values of the bosonic fields which are different from the usual Lorentzian ones: these can be imposed 
on on-shell fields (e.g.~$A_0$ or~$g_{0i}$ are typically imaginary as dictated by the Wick rotation), or 
on off-shell BPS fluctuations (e.g.~the localization manifold on~AdS$_2$ involves scalar fields 
with~$\bar{X} \neq X^*$ \cite{Dabholkar:2010uh, Gupta:2012cy,Gupta:2015gga}.}

\medskip

\ndt \textbf{Supermultiplets:} 
For the $\mathcal{N}=1$ 5d conformal supergravity theory, we follow the conventions of \cite{deWit:2009de}.
We consider the Weyl multiplet, which couples to $N_\text{v}$ number of vector multiplets as well as a single hyper multiplet.  
Independent fields in each multiplet is summarized in Table~\ref{table:fieldcontent}.

The Weyl multiplet consists of the gauge fields corresponding to all the symmetry generators 
of~$\cN= 1$ superconformal algebra  $\{ P^A\,, M^{AB}\,, D\,,K^A\,, Q^i\,,S^i \,,V_{j}{}^{i} \}$. 
Among all the gauge fields, the gauge fields associated with $\{ M^{AB}\,,K^A\,, S^i  \}$ are 
composite, i.e.  they are expressed in terms of other gauge fields. The independent gauge fields 
in Weyl multiplet are the vielbein $E_M{}^A$, dilatation gauge field $b_M$,  gaugino $\psi_M^i$, 
and the $SU(2)_R$ gauge field $V_{M j}{}^i$. For the Weyl multiplet to be off-shell supermultiplet, 
it includes  auxiliary two-form tensor $T_{AB}$, auxiliary fermion $\chi^i$, and auxiliary scalar $D$. 
Hence the independent fields of the Weyl multiplet are summarized as
\be
\mbox{Weyl:~~~}\{ {E_M}^A \, ,\,\Psi^i_M \, ,\, b_M \, ,\, {V_{M,\,i}}^j \, ;\,T_{MN}\, ,\,\chi^i\,,\, D \}\,.
\ee
Here, the indices $\{A, B, \cdots\}$,  $\{M, N, \cdots\}$,  $\{i, j, \cdots\}$ are five-dimensional flat tangent space,  curved spacetime, and~$SU(2)$ fundamental indices, respectively, which are summarized in appendix \ref{sec:spinors}. In the Weyl multiplet fields, we use the special conformal symmetry (that acts only on~$b_M$) to gauge-fix~$b_M=0$, 
so that from here on this field will not appear.
We consider $N_\text{v}$ vector multiplets labeled by $I$, each of which consists of
\be
\mbox{Vector:~~~}\{ \sigma^I \, ,\,W^I_M \, ,\, \Omega^{I\,i} \, ,\, Y^I_{ij}\}\,,\qquad I = 1\,, 2\,, \cdots \,,N_\text{v}\,.
\ee
They corresponds to  a scalar, a~$U(1)$ gauge field, gaugini, and 
an auxiliary symmetric~$SU(2)$ triplet.
The $i,j$ indices are raised and lowered using the~$SU(2)$ 
symplectic metric~$\varepsilon$, where, explicitely,~$\varepsilon_{12} = \varepsilon^{12} = 1$.
In particular, we have~$Y_{ij} \= \varepsilon_{ik} \, \varepsilon_{j \ell} \, Y^{k \ell}$.
We finally consider a single  hypermultiplet, which  consists 
of 
\be
\mbox{Hyper:~~~}\{ A_i{}^\alpha \, ,\,\zeta^\alpha\}, 
\ee
corresponding to the hyper scalar,  and the hyper fermion, where~$\alpha = 1,2$. 
There is no known off-shell Lorentz-covariant hypermultiplet with finite number of fields. 
In the backgrounds that we consider below, the hypermultiplet turns out to take its on-shell value, 
and we can therefore integrate it out at the semi-classical level. For the full localization calculation, 
one could construct an off-shell hypermultiplet for one complex supercharge (see e.g.~\cite{Hama:2012bg,Murthy:2015yfa}).  
One of the~$N_\text{v}$ vector multiplets and the single hypermultiplet act as the two compensators 
to be added to the Weyl multiplet to form a 5d $\cN=1$ Poincar\'e supergravity 
multiplet.

\vskip 0.2cm

\ndt \textbf{Supersymmetry algebra:}
The supersymmetry transformations of the various spinor fields under the~$Q$ and~$S$ supersymmetry
transformations are given in~\eqref{KSequation1}. 
Two $Q$-supersymmetry transformations, parametrized by spinors~$\epsilon_1$ 
and~$\epsilon_2$ respectively, close onto the bosonic symmetries of the theory as
\be\label{QQalgebra}
[ \delta_Q{(\epsilon_1)}\,, \delta_Q{(\epsilon_2)} ] \= 
\delta_{\text{gct}}(\xi^\mu) +\delta_M(\lambda) +\delta_S(\eta)+\delta_K(\Lambda_K)
\ee
where~$\delta_{\text{gct}}$ are the general coordinate transformations,~$\delta_M$ is a local 
Lorentz transformation, $\delta_S$ is a conformal supersymmetry transformation, and $\delta_K$ 
is special conformal transformation.
The relevant parameters to this paper are 
\be
\begin{split}
\xi^\mu & \= 2 \bar{\epsilon}_{2i} \gamma^\mu \epsilon^{i}_1\,,\\
\lambda^{AB} & \= -\xi^\mu \omega_\mu^{~AB} +\frac{\i}{2} T^{CD} \bar{\epsilon}_{2i} 
(6 \gamma^{[A} \gamma_{CD}\gamma^{B]} -\gamma^{AB}\gamma_{CD}-\gamma_{CD}\gamma^{AB}  )\epsilon^i_1\,.
\end{split}
\ee

\vskip 0.2cm

\ndt \textbf{Action:} 
\ndt The bosonic Lagrangian at two-derivative level is
\begin{equation} \label{action}
\begin{split}
    L_{\text{bulk}}  \= E\,\left(\mathcal{L}_V + \mathcal{L}_{VW} + \mathcal{L}_H+\mathcal{L}_{HW}+\mathcal{L}_{CS}\right)\,,
\end{split}
\end{equation}
where~$E \equiv \text{det}({E_M}^A)$,~$\mathcal{L}_V$ contains purely vector multiplet terms,~$\mathcal{L}_{VW}$ 
contains mixing between vector and Weyl,~$\mathcal{L}_{H}$ is the kinetic hyper scalar piece,~$\mathcal{L}_{HW}$ 
contains coupling of hyper to Weyl, and~$\mathcal{L}_{CS}$ is the five-dimensional Chern-Simons action:
\begin{equation} \label{action}
\begin{split}
     \mathcal{L}_V & \= \frac{1}{2}\,c_{IJK}\,\sigma^I  \,\left(\frac{1}{2}\,D^M\,\sigma^J\,D_M\,\sigma^K
     +\frac{1}{4}\,F^J_{MN}\,F^{MN\, K}-3\sigma^J\,F^K_{MN}\,T^{MN}-Y_{ij}^J\,Y^{K\,ij}\right)\,, \\
     \mathcal{L}_{VW} & \= -C(\sigma)\left(\frac{1}{8}\,R-4D-\frac{39}{2}\,T^2\right)\,,\\
     \mathcal{L}_H &\=-\frac{1}{2}\Omega_{\alpha \beta}\varepsilon^{ij}D_M\,{A_i}^\alpha \, D^M\,{A_j}^\beta\,,\\
     \mathcal{L}_{HW} &\= \chi\,\left(\frac{3}{16}\,R+2D+\frac{3}{4}\,T^2\right)\,,\\
     \mathcal{L}_{CS} &\=-\frac{i}{48\,E}\varepsilon^{MNOPQ} c_{IJK} W^I_M F^J_{NO}F^K_{PQ} \,.
\end{split}
\end{equation}
In the Chern-Simons Lagrangian~$\mathcal{L}_{CS}$, the object~$\varepsilon^{MNOPQ}$ is a fully 
antisymmetric tensor density taking values in~$\{-1,\,0,\,1\}$.
The scalar norms appearing in $\mathcal{L}_{VW}$ and $\mathcal{L}_{HW}$ are:
\begin{align}\label{Csigma}
& C(\sigma) \defeq \frac{1}{6\,}c_{IJK}\,\sigma^I \sigma^J \sigma^K\,,\\
& \chi \defeq \frac{1}{2}\,\Omega_{\alpha \beta}\, \varepsilon^{ij} {A_i}^\alpha {A_j}^\beta\,.
\end{align}
The action of the theory is 
\be \label{bulk_action}
S_{\text{bulk}} \= \frac{1}{8\pi^2}\int _{\mathcal{M}} d^5x \,L_{\text{bulk}}\,,
\ee
for an appropriate  coordinate chart on the 5d manifold $\mathcal{M}$.

\subsection{Global Lorentzian AdS$_3 \times$ S$^2$} \label{lorentzian}

We consider the fully supersymmetric {AdS$_3 \times$ S$^2$ solution of the Lorentzian supergravity described above, 
corresponding to the near-horizon geometry of the half-BPS magnetic black string \cite{Elvang:2004rt}. The  metric in Lorentzian signature is  
\be \label{metric}
ds^2  \=4 \ell^2\,(-\cosh^2\rho\,dt^2+d\rho^2+\sinh^2 \rho \, d\psi^2)+\ell^2\,(d\theta^2+\sin^2 \, \theta d\phi^2)\,,
\ee
where the coordinates of the AdS$_3$ have the ranges~$\rho \in [0, \, \infty),\, \psi \in [0, \, 2\pi] ,\,  t \in (- \infty,\, \infty)$
and the angles on the S$^2$ have ranges~$\theta \in [0,\,\pi] ,\, \phi \in [0,\, 2\pi]$.
The radii of the~AdS$_3$ and the~S$^2$ are~$(2\ell)$ and~$\ell$ respectively,
where this relative factor of~$2$ is determined by supersymmetry.
Note that in the off-shell theory,~$\ell$ is free and parametrizes the dilatations of the theory,
while in the on-shell theory (where dilatations are broken) it is determined by the magnetic charges 
of the solution via the D-gauge condition.
These magnetic charges~$p^I$ enter the solution through the vector multiplet. 
The non-trivial fields of the vector multiplet are: 
\be \label{vector_field_content}
\sigma^I \= -\frac{ p^I}{\ell} \, , \qquad  F^I_{\theta \phi} \= p^I\,\sin \theta\,.
\ee
Note that the solution does not have electric flux, which allows us to turn on flat gauge connections on the~AdS$_3$.
This aspect will become relevant in the following.
The five-dimensional Newton's constant is~$G_5 = 6 \pi \ell^3/p^3$ and the three-dimensional
Newton's constant is obtained by setting~$G_5 = \text{Area}_{S^2} \times G_3 = 4\pi \ell^2 G_3$.

In the off-shell formalism of Section~\ref{5dSUGRA}, one requires additional auxiliary fields. 
In the Weyl multiplet, the non-trivial fields are:
\be \label{TMN}
     T_{\theta \phi}  \=- \frac{\ell}{4} \, \sin \theta\,.
\ee
In the compensating hypermultiplet, the BPS equation is solved by 
\begin{align} \label{hyper_field_content}
{A_i}^\alpha \= {c_i}^\alpha\,, \quad  
\end{align}
where the constants ${c_i}^\alpha $ are determined in terms of the charge $p^I$ by the field equation for the auxiliary field $D$ to be 
\be \label{cprelation}
\Omega_{\alpha\beta}\,\varepsilon^{ij}c_i{}^\alpha c_j{}^{\beta} \= \frac{2}{3 \ell^3}\, c_{IJK}\,p^I p^J p^K\,.
\ee
In this paper, we fix an explicit choice for the~$c_i{}^\alpha$ as
\be
c_1{}^2\=c_2{}^1\=0\,, \qquad c_1{}^1\=c_2{}^2\= \sqrt{\frac{p^3}{3\ell^3}}\,.
\ee

\subsection{Supersymmetry algebra in Lorentzian AdS$_3 \times$ S$^2$}\label{LorentzianKSandAlgebra}

\ndt\textbf{Killing spinors} The $Q$- and $S$- supersymmetry parameters,~$ \epsilon^i$ and~$\eta^i$ 
respectively, that are preserved by the bosonic fields of the global AdS$_3\times S^2$ background are 
determined by setting the variation of gravitino and the variation the auxiliary fermion in~\eqref{KSequation1} to zero.
These two equations are, respectively, 
\beqa \label{KSE}
&&0\= 2 \, \mathcal{D}_M \epsilon^i +\frac{\i}{2}T_{AB}(3\gamma^{AB}\gamma_M-\gamma_M\gamma^{AB})
\epsilon^i-\i \gamma_M \eta^i \,.
\\
\label{auxiliaryKSE}
&&0 \= \frac{1}{2}  \epsilon^i D +\frac{1}{64} R_{MN j}{}^i(V) \gamma^{MN}\epsilon^j 
+\frac{3}{64}\i (3\gamma^{AB} \slashed{D} +\slashed{D}\gamma^{AB} )
T_{AB}\epsilon^i \\
&&~~~~~ \quad -\frac{3}{16}T_{AB}T_{CD}\gamma^{ABCD}\epsilon^i +\frac{3}{16}T_{AB}\gamma^{AB}\eta^i\,.  \nn
\eeqa
On our bosonic background, the second equation \eqref{auxiliaryKSE} immediately determines 
the $S$-supersymmetry spinor as $\eta^i=0$. 
The first equation is referred to as the Killing spinor equation. We analyze its solutions in 
Appendix~\ref{KSads3s2} and summarize the results below.

The complex basis of the Killing spinor on AdS$_3 \times$ S$^2$ is given by the 
the following four Killing spinors, 
\be\ba{ll}\label{4complexKS}
\epsilon_+^{~+}= \sqrt{\frac{\ell}{2}}\,\epsilon^+_{\text{AdS}_3}\otimes \epsilon^{+}_{\text{S}^2} \,, 
\quad\quad\quad&\epsilon_+^{~-}= \sqrt{\frac{\ell}{2}}\,\epsilon^+_{\text{AdS}_3}\otimes \epsilon^{-}_{\text{S}^2} \,,\\
\epsilon_-^{~+}=\sqrt{\frac{\ell}{2}}\, \epsilon^-_{\text{AdS}_3}\otimes \epsilon^{+}_{\text{S}^2} \,, 
\quad\quad\quad&\epsilon_-^{~-}= \sqrt{\frac{\ell}{2}}\,\epsilon^-_{\text{AdS}_3}\otimes \epsilon^{-}_{\text{S}^2} \,,
\ea\ee
with
\be\label{KSads3pm}
\begin{split}
\epsilon^{+}_{\text{AdS}_3}& \= {\rm e}^{\frac{\i}{2}(t +\psi)}\left( \ba{c}\cosh \frac{\rho}{2} \\ -\sinh\frac{\rho}{2}\ea \right)\,, 
\qquad  \epsilon^{-}_{\text{AdS}_3}\= {\rm e}^{-\frac{\i}{2}(t +\psi)}
\left( \ba{c}-\sinh \frac{\rho}{2} \\ \cosh\frac{\rho}{2}\ea\right)\,, \\
\epsilon^{+}_{\text{S}^2} & \= {\rm e}^{\frac{\i}{2}\phi} \left( \ba{c}\cos \frac{\theta}{2} \\ \sin\frac{\theta}{2}\ea \right)\,, 
\qquad\qquad\qquad 
\epsilon^{-}_{\text{S}^2}\= {\rm e}^{-\frac{\i}{2}\phi} \left( \ba{c}-\sin \frac{\theta}{2} \\ \cos\frac{\theta}{2}\ea \right)\,.
\end{split}
\ee
These 4 Killing spinors organize themselves into the 8 pairs of symplectic Majorana spinors 
\be\label{8SMKS}
\begin{split}
&\epsilon^{\,i}_{(1)}=(-\i \epsilon_+^{~+}, \epsilon_-^{~-})\,, \quad \; \epsilon^{\,i}_{(2)}=(\epsilon_+^{~+},-\i\epsilon_-^{~-})\,, \quad \;
\epsilon^{\,i}_{(3)}=\, -(\epsilon_-^{~-},\i\epsilon_+^{~+})\,, \quad  \; \epsilon^{\,i}_{(4)}=\, -(\i\epsilon_-^{~-},\epsilon_+^{~+})\,,\\
&\tilde{\epsilon}^{\,i}_{(1)}\=(\epsilon_+^{~-},  \i\epsilon_-^{~+})\,,\quad \; 
\tilde\epsilon^{\,i}_{(2)}\=(\i\epsilon_+^{~-},\epsilon_-^{~+})\,,\quad \;\,
\tilde\epsilon^{\,i}_{(3)}\=(-\i\epsilon_-^{~+}\,,\epsilon_+^{~-})\,,\quad \;
\tilde\epsilon^{\,i}_{(4)}\=(\epsilon_-^{~+}\,,-\i\epsilon_+^{~-})\,,
\end{split}
\ee
to form the 8 real basis of the Killing spinor on AdS$_3 \times $S$^2$. 
Each pair satisfies the following symplectic Majorana condition~\eqref{SMLor} appropriate to the 5d Lorentzian theory, 
i.e.~$(\epsilon^{i})^{\dagger}\gamma_{\hat t}\= \varepsilon_{ij} ( \epsilon^j)^T \cC$ in the conventions of Appendix~\ref{KSads3s2}.

\bigskip

\ndt \textbf{Superconformal algebra:} 
 Let us denote  
 \be
 \cQ_a \= \delta(\epsilon^i_{(a)})\,,~~~~~~~~~~\wt\cQ_{a} \= \delta(\tilde\epsilon^i_{({a})})\,,~~~~a= 1\,,2\,,3\,,4\,,
  \ee
 with the Grassmann even Killing spinors~$\epsilon^i$. 
 Then,
\beqa
&& \bigl\{ \cQ_a\,, \cQ_b \bigr\} \=
- 2 \i   \delta_{ab}  (L_0 - J^3)\,, \qquad  \bigl\{ \wt\cQ_a\,, \wt\cQ_b \bigr\} \=- 2 \i   \delta_{ab}  (L_0 + J^3)\,, \\
&&
\bigl\{ \cQ_a\,, \wt\cQ_b \bigr\} \= 
\begin{pmatrix} -2\i J^2 & 2\i J^1& -(L_+ -L_-)& \i (L_+ + \L_-)\\- 2\i J^1& -2\i J^2  
	& -\i (L_+ +L_-)&-(L_+ -L_-)  \\L_+ -L_- & \i (L_+ + L_-)& -2\i J^2 & -2\i J^1 \\ -\i (L_+ +L_-) 
	& L_+ - L_- & 2\i J^1 &- 2\i J^2\end{pmatrix}\,,
\eeqa
where $SL(2,R)$ generators $L_0\,, L_\pm $ and  $SO(3)$ generators  $ J^\boa$ satisfy
\be \label{L0LpmJi}
[L_+,\,L_-] \= -2 L_0\,,\qquad[L_0\,,L_\pm]\= \pm L_\pm\,,\qquad[J^\boa \,,J^\bob]\= \i \epsilon^{\boa\bob\boc}J^\boc\,.
\ee
Their explicit representation on the~$\adss$ is given in Appendix~\ref{isogenerator}.

\bigskip

Let us define the supercharges $G^{i\alpha}_\gamma$ 
\be \label{n=4supercharges2}
\begin{split}
& G_+^{++} \, \equiv \, \frac{\i \cQ _{1} +\cQ _{2}}{2} \,, \quad 
G_-^{+-} \, \equiv \,  \frac{- \cQ _{3}+\i \cQ _{4}}{2}\,, \quad 
G_+^{+-}  \, \equiv  \,  \frac{\wt\cQ _{1}-\i \wt\cQ _{2}}{2} \,, \quad \; \,
G_-^{++} \, \equiv \, \frac{\i \wt\cQ _{3} + \wt\cQ _{4}}{2}  \,, \\ 
& G_-^{--} \, \equiv \, \frac{\cQ _{1} + \i \cQ _{2}}{2} \,, \quad 
G_+^{-+} \, \equiv \, \frac{\i  \cQ _{3} -\cQ _{4}}{2} \,, \quad \; \;
G_-^{-+} \, \equiv \, \frac{-\i \wt\cQ _{1} +\wt\cQ _{2}}{2} \,, \quad 
G_+^{--} \, \equiv  \,  \frac{\wt\cQ _{3} +\i \wt\cQ _{4}}{2} \,,
\end{split}
\ee
where $\gamma$ is the sign of the~$L_0$ eigenvalue, $i$ is the outer automorphism from 
the~SU(2) R-symmetry of the supergravity, and $\alpha$ is the~SU(2) R-symmetry index corresponding to 
isometries of the~S$^2$.
Then, we obtain the non-trivial commutation relations:
\be \label{Gpmanticom}
\bigl\{G^{+\alpha}_{\pm}\,,G^{- \beta}_{\mp} \bigr\} \=  \epsilon^{\alpha \beta}L_0 \pm  (\epsilon \btau_\boa)^{\beta\alpha} J^\boa\,,
\qquad \bigl\{ G^{+\alpha}_\pm \,, G^{-\beta}_\pm \bigr\}\= \mp  \i \epsilon^{\alpha \beta} L_\pm\,,
\ee
and  
  \be \label{LpmGcom} 
  \ba{ll}
 \left[L_0 \,, G^{i \alpha}_{\pm} \right] \= \pm \half \,G^{i\alpha}_{\pm}\,,~~~~~ &
 \left[L_\pm\,, G^{i\alpha}_{\mp}\right] \= -\i \,G^{i\alpha}_{\pm}\,, \\
 \left[J^3 \,, G^{i \pm}_{\gamma}\right] \=  \pm \half \,G^{i \pm}_{\gamma}\,,~~~
 &\left[J^\pm \,, G^{i \mp}_{\gamma}\right] \=   \,G^{i \pm}_{\gamma}\,,
\ea\ee
where  $J^\pm \equiv J^1 \pm \i J^2$. 
The algebra~\eqref{L0LpmJi},~\eqref{Gpmanticom},~\eqref{LpmGcom} is 
the global part~$su(1,1|2)$ of the NS-sector chiral~$\cN=4$ superconformal algebra. 
Denoting the super Virasoro charges as~$\cL_n$, $n \in \IZ$ and~$\cG^\alpha_{\dot A,r}$, 
$r \in \IZ+\half$,~$\dot A =(+,-)$, the embedding into the ~$\cN=4$ superconformal algebra as presented 
e.g.~in~\cite{Guo:2022ifr} is given by
{${L_{\pm} = \mp \i {\cL}_{\mp 1}}$}, $L_0=\cL_0$,
$G^{\pm \alpha}_{\pm}  = \pm \,{\cG}^{\alpha}_{\mp, \mp 1/2}$,
 $G^{\mp \alpha}_{\pm}  = \pm  \, {\cG}^{ \alpha}_{\pm, \mp 1/2}$,
 and the $su(2)$ zero-modes are unchanged.

\section{Supersymmetric $\Hthrees$ and twisting}  \label{EuclBackground}

In this section we move from the Lorentzian $\adss$ configuration to the Euclidean $\Hthrees$ geometry. 
We begin, in the first subsection, by reviewing the thermal~AdS$_3$ (=$\IH^3/\IZ$) geometry and some aspects 
of the AdS/CFT correspondence in this set up. 
In the second subsection we move to the supersymmetric version of the thermal geometry which 
requires a non-trivial twist. In the third subsection we discuss the Hamiltonian trace interpretation of 
the functional integral on this twisted configuration, and discuss how this is related to the elliptic genus in 
the semi-classical limit.

\subsection{Thermal compactification of AdS$_3$ \label{adscft} }

In this subsection we review the general set-up of the~AdS$_3$/CFT$_2$ dictionary in the 
context of our problem, following the treatment of~\cite{Kraus:2006wn}.
We begin with the three-dimensional pure Einstein-Hilbert action with a cosmological constant: 
\be \label{EH3d}
S_{\text{grav}} \= \frac{1}{16\pi G_3}\int d^3x\sqrt{g}(R-\frac{1}{2\ell^2})\,.
\ee
AdS$_3$ is a solution of the equations of motion of this theory with constant negative curvature. 
The Wick rotation~$t = -\i t_E$ 
leads to the Euclidean~($\IH^3$) metric 
\be \label{ads3metric}
ds^2 \= 4\ell^2 \left(\cosh^2 \rho \, {dt_E}^2 +  d \rho^2 + \sinh^2 \rho \, d \psi^2 \right)\,.
\ee
When the range of~$t_E$ is~$(-\infty, \infty)$ we obtain the cylinder geometry corresponding to the global 
solution. 
Thermal AdS$_3$ corresponds to the quotient~$\IH^3/\IZ$, obtained by imposing the periodicities
\be \label{torusperiodicities}
\bigl(t_{\scriptscriptstyle{E}} \,, \psi  \bigr) \sim \bigl(t_{\scriptscriptstyle{E}} + 2\pi \tau_2\,, 
\psi + 2\pi \tau_1 \bigr) \sim \bigl(t_{\scriptscriptstyle{E}}\,, \psi + 2\pi  \bigr)\,,
\ee
so that the geometry is that of a solid torus. 
Physically, this corresponds to setting the chemical potential conjugate to angular momentum 
(i.e.~the angular velocity) to~$2 \pi \tau_1$ and the chemical potential conjugate to energy (i.e.~the inverse temperature) 
to~$2 \pi \tau_2$.

It is convenient to introduce the coordinates $z \equiv \psi + \i \, t_E$, $\bar{z} \equiv \psi\ - \i \,t_E$. 
Upon taking the large-$\rho$ expansion of the transverse metric 
tensor~${g}_{\alpha \beta}$, where~$x^\alpha = (z,\,\bar{z})$,
one obtains the Fefferman-Graham form
\be \label{FGmetric}
{g}_{\alpha \beta} \= \rme^{2\rho}{g}^{(0)}_{\alpha \beta} + {g}^{(2)}_{\alpha \beta} +\cdots\,.
\ee
The conformal boundary metric~${g}^{(0)}_{\alpha \beta}$ is
\be \label{flatT2}
{g}^{(0)}_{\alpha \beta} \= \ell^2\, {dz\,d\bar{z}}\,,
\ee
with the identifications
\be \label{zzbarid}
(z,\bar{z}) \sim (z + 2\pi, \bar{z} + 2 \pi) \sim  (z+2\pi\tau , \bar{z} + 2 \pi \bar{\t}) \,,
\ee
consistent with the interpretation that the boundary CFT lives on the flat torus with modular parameter~$\tau$.

\vskip 0.2cm

\ndt \textbf{Boundary conditions on the gauge fields}\\
More generally, we include constant chemical potentials~$\mu^I$ for a number of conserved~$U(1)$ 
charges~$q_I = \int J_I$  where~$J_I$ are the corresponding conserved currents in the boundary CFT. 
The partition function of such a CFT is
\be \label{trace}
\text{Tr}_{\mathcal{H}_\text{CFT}}\, \rme^{- \beta H  +  \mu^I q_I}\,. 
\ee
The dual gravitational theory~\eqref{EH3d} includes the same number of~$U(1)$ 
gauge fields\footnote{In this subsection,~$W^I$ and all other fields are three-dimensional.}~${W}^I$.
The most relevant term governing their dynamics at low energies is given by the Chern-Simons action 
\be \label{CS3d}
 -\frac{\i}{8 \pi}k_{IJ} \int  W^I \wedge dW^J \= -\frac{\i}{8 \pi}k_{IJ} \int d^3x \, 
 \varepsilon^{\mu \nu \lambda}\,W^I_\mu \, \partial_\nu W^J_\lambda \,.
\ee
(Here, unlike in the rest of the paper, we
employ the indices~$\mu,\,\nu\,\cdots$ to denote the 3d coordinates~${x^\mu=(\rho,\,x^\alpha)}$.)
In the gauge~$W^I_\rho = 0$, the gauge fields admit a large-$\rho$ expansion analogous to~\eqref{FGmetric} as:
\be \label{FGexpQ}
{W}^I_\alpha \= {W}^{I\,(0)}_\alpha+\rme^{-2 \rho}\, {W}^{I\,(2)}_\alpha+\cdots\,.
\ee
The asymptotic equations of motion imply that~${W}^{I\,(0)}_\alpha$ is flat.

As is well-known, the fact that the CS term has a first order kinetic term so that the two legs~${W}^{I}_{z,\,\bar{z}}$ 
form canonical pairs in the Hamiltonian theory \cite{Elitzur:1989nr}. 
One should therefore impose Dirichlet boundary conditions on only one of the legs:
\be \label{CSbc}
\delta {W}^{I \, (0)}_z\=0 \,, \qquad {W}^{I\,(0)}_{\bar{z}} \; \text{not fixed.}
\ee
Now, in accord with the bulk/boundary correspondence, the boundary source~$\mu^I$
must be identified with the asymptotic value of the gauge field~${W}^{I\,(0)}_z$. 
Since the~$\psi$-cycle is contractible, any smooth configuration must have~$W^I_\psi=0$ at the origin.
The saddle-point configurations have flat gauge fields due to the equations of motion,  
and therefore obey 
\be \label{Wzzbar}
{W}^{I}_z  \=  -W^I_{\overline{z}} \= -\i\, \mu^I \,.
\ee

The AdS/CFT correspondence states that the trace \eqref{trace} equals 
the following bulk functional integral, up to a Casimir-energy-like term, 
\be \label{PI3d}
Z_\text{AdS}(\tau,\,\mu) \= \int D \phi_\text{grav} \; \rme^{-S_{\text{ren}}(\tau,\,\mu)} \,,
\ee
where~$\phi_\text{grav}$ denotes the gravitational fields of the theory, and 
\be
S_{\text{ren}}\,  \equiv \,  S_{\text{bulk}}+S_{\text{bdry}}
\ee
is the renormalized action of the gravitational theory.
Here,~$S_{\text{bulk}}$ is the bulk Euclidean action of the Einstein-Hilbert-matter theory 
and~$S_{\text{bdry}}$ is the boundary action required to make the total action finite and well-defined 
under our choice of boundary conditions.
In particular, it includes the Gibbons-Hawking boundary term, and a Chern-Simons boundary term given by 
\be \label{CSbdry}
- \frac{\i}{8 \pi}\, k_{IJ}\, \int dz \, d\bar{z} \, \Big[{W}^I_z \, {W}^J_{\bar{z}}\Big]_{\text{bdry}}\,.
\ee
This last term is required to ensure the consistency of the variational principle of the gauge fields with the 
boundary conditions~\eqref{CSbc}. 

\vskip 0.2cm
\ndt
\textbf{The semi-classical contribution:}
At leading order, the partition function~\eqref{PI3d} is given by the value of~$S_{\text{ren}}$ on 
the thermal AdS$_3$ configuration described above. 
The value of this action is~\cite{Kraus:2006wn} 
\be \label{thermalaction}
S_{\text{ren}}(\tau ,\, \mu) \= -\pi  \tau_2 k  - \pi \tau_2 \,k_{IJ} \, \mu^I \mu^J\,,
\ee
where~$6k = \frac{3 (2\ell)}{2 G_3}$ is the Brown-Henneaux central charge of the gravitational theory for 
the AdS$_3$ space~\eqref{ads3metric}, and~$k_{IJ}$ is the level of the Chern-Simons term~\eqref{CS3d}. 
Here, note that the boundary~$U(1)$ current obtained from~\eqref{CS3d},~\eqref{CSbdry} is right-moving.
The choice of opposite relative sign between~\eqref{CS3d} and~\eqref{CSbdry} leads to the opposite chirality.

In the context of the five-dimensional theory of the previous section, recall from the discussion 
around~\eqref{vector_field_content} that~$G_5 = 6 \pi \ell^3/p^3$, so that 
\be \label{cgravval}
6 k \= \frac{3 (2\ell)}{2 G_3} \= \frac{12 \pi \ell^3}{G_5} \= 2 p^3\,.
\ee
To identify~$k_{IJ}$, we reduce the five-dimensional Chern-Simons boundary term in~\eqref{CSbdr} 
onto the S$^2$, and compare the resulting action with the three-dimensional Chern-Simons boundary 
term~\eqref{CSbdry} on thermal AdS$_3$.
This leads to:
\be \label{kIJ}
 k_{IJ} \= \frac{2}{3}c_{IJK}\,p^K\,.
\ee

\subsection{Twisted background and superalgebra \label{sec:twistedbackgnd}}

Now we consider the supersymmetric theory on the 5d geometry 
Under the Wick rotation~$t = - \i t_E$, the Lorentzian metric~\eqref{metric} rotates to that of 
Euclidean $\IH^3 \times S^2$.
The non-trivial fields in the Weyl multiplet are:
\begin{eqnarray} \label{Emetric}
ds^2  &\= & 4\ell^2\,(\cosh^2\rho\,dt_{{E}}^2+d\rho^2+\sinh^2\rho \,d\psi^2)
+\ell^2\,(d\theta^2+\sin^2\theta\, d\phi^2)\,,\\
T_{\theta \phi}  & \= & - \frac{\ell}{4} \, \sin \theta\,.
\label{ETMN}
\end{eqnarray}
If the Euclidean time coordinate~$t_E$ runs from~$(-\infty,\, \infty)$, 
the topology is that of a solid cylinder times a sphere, which we call the Euclidean cylinder frame.
Although the Killing spinor equations~\eqref{KSE} and~\eqref{auxiliaryKSE} are formally solved by the 
same set of eight spinors~\eqref{8SMKS} in this background,
these spinors are no longer well-defined because they diverge at the ends of the Euclidean cylinder. 
The solution to this problem involves compactifying the Euclidean time on a circle and 
simultaneously rotating the~S$^2$ as we go around the time circle.
This twisted quotient makes for a well-defined background, as we now describe.

We start from the configuration~\eqref{Emetric} describing an infinite solid cylinder (times a sphere),
and make the following identifications, 
\be  \label{twistwithtau1}
( t_E\,, \psi \,, \phi)  \; \sim \; ( t_E \,, \psi +2\pi \,, \phi  )
\; \sim \; ( t_E + 2\pi \tau_2\,, \psi +2\pi \tau_1\,, \phi  + \i 2 \pi \tau_2  \Omega ) \,.
\ee
Equivalently, we can define a  new  set of ``twisted" coordinates, 
\be\label{twistedcoord}
t'_E \= t_E\,,\qquad \phi' \; \equiv  \; \phi -\i \Omega t_E\,,
\ee
which have the identification 
\be \label{Newperiodicity}
( t_E'\,, \psi \,, \phi') \; \sim \; ( t_E' \,, \psi+2\pi\,, \phi'  ) \; \sim \; ( t_E' + 2\pi \tau_2\,, \psi+2\pi\tau_1\,, \phi' ) \,.
\ee
We denote the corresponding complex coordinates 
as~$z'=\psi+\i t_E'\,$, $\bar{z}' = \psi-\i t_E'\,$, identified as~$( z'\,, \bar{z}')  \; \sim \; ( z' + 2\pi \tau\,, \bar{z}'+2\pi\bar{\tau})$.

In the twisted frame, the on-shell background configuration is 
\be \label{TwistedFrame}
\begin{split}  
ds^2 & \=4 \R^2 \left(\cosh^2\! \rho \, dt'^2_E + d\rho^2  + \sinh^2 \!\rho \, d\psi^2 \right) + 
\R^2 \Bigl(d\theta^2 + \sin^2\theta  \bigl(d\phi' + \i \Omega dt'_E\bigr)^2 \Bigr)\,,\\
T_{\theta \phi'}  & \=  - \frac{\ell}{4} \, \sin \theta\,,\qquad T_{\theta t_{E}'} \=  -\i  \frac{\ell }{4} \Omega\, \sin \theta\,, \\
\sigma^I &\= -\frac{p^I}{\ell} \,, \qquad  
W^I_{{t_E}'} \= 2 \mu^I -\i \Omega \,p^I \cos{\theta}\,, \qquad W^I_{\phi'} \= -p^I \cos \theta \,,
\\
A_1{}^1 & \=A_2{}^2\= \sqrt{\frac{p^3}{3\ell^3}}\,.
\end{split}
\ee
The~S$^2$ in~\eqref{TwistedFrame} is fibered over the time circle of AdS$_3$, and 
we refer to this configuration as the \textit{twisted torus} background.
We also note that in the expression for~$W^I_{{t_E}'}$ we have introduced an arbitrary 
constant
$\mu^I$ which is allowed by the supersymmetry and equations of motion, 
which we will interpret as the source of a $U(1)$ current in the boundary CFT.
The BPS equations may also allow~$W^I_{\psi}$ to take a constant value,
but this constant is forced to be zero due to the contractibility of the~$\psi$-cycle.

To see that the twisted torus background \eqref{TwistedFrame} has well-defined supersymmetry, 
we solve the Killing spinor equation from the variation of gravitino \eqref{KSequation1}, which is rewritten 
now as
\be\label{TwistKSE}
0\= 2\cD_M \varepsilon^i -\frac{\i }{4\ell} 
(3 \gamma^{\hat\theta \hat\phi}\gamma_M - \gamma_M \gamma^{\hat\theta \hat\phi}) \, \varepsilon^i\,.
\ee
Here we use the following gamma matrices in the Euclidean theory, 
which follow from the Wick rotation, 
\be \label{gamma5dE}
  \gamma_{\hat{t}_E}= \bsigma_3 \otimes \btau_3\,,  
    \quad \gamma_{\hat\rho}= 
    \bsigma_1 \otimes \btau_3\,,  
     \quad \gamma_{\hat\psi}=
   \bsigma_2 \otimes \btau_3 \,,  
    \quad  \gamma_{\hat{\theta}}=
        \mathbb{I} \otimes \btau_1\,,
   \quad \gamma_{\hat\phi}=\mathbb{I} \otimes \btau_2\,, 
\ee
where we relate the $\sigma_3$ with the Lorentzian gamma matrix $\bsigma_0$~in \eqref{gamma5dLor} 
by $\bsigma_3\equiv -\i\bsigma_0 $. We will take the representation 
$(\bsigma_3\,,\bsigma_1\,,\bsigma_2) =(-\btau_3\,,\btau_1\,,\btau_2)$ with the Pauli sigma matrix $\btau_a$. 
Note that unlike the case of global AdS$_3 \times $ S$^2$ in the subsection \ref{LorentzianKSandAlgebra}, 
the Killing spinor equation \eqref{TwistKSE} does not split into the equations of AdS$_3$ and S$^2$.
This is because we have the following spin connections 
\be\label{SpinconnectionTwist}
\omega_{t'_E}^{12}= -\sinh\rho\,, ~~~\omega_{t'_E}^{45}=\i \Omega \cos\theta\,, ~~~\omega_{\psi}^{23}= \cosh\rho\,,~~~\omega_{\phi'}^{45}= \cos\theta \,,
\ee
where there is mixing between AdS$_3$ and S$^2$ directions through the non-zero twisting parameter $\Omega$. 

The solution of Killing spinors can be easily found by following the twisting construction. 
It is clear that the Euclidean continuation of the set of 8 Lorentzian Killing spinors~\eqref{4complexKS}, \eqref{8SMKS}, 
followed by the coordinate transformation~\eqref{twistedcoord} obeys the new Killing spinor equation.  
Upon setting the parameter 
\be \label{OmegaValue}
\Omega \= 1+ \i \frac{\tau_1}{\tau_2}\,,
\ee
the following\footnote{The choice $\Omega= -1 -\i \frac{\tau_1}{\tau_2}$ 
also gives rise to a different 4 set of Killing spinors.} 4 of the original~8 Killing spinors 
\be \label{4TwistedSMKS}
\varepsilon^{\,i}_{(1)}=(-\i \varepsilon_+^{~+}\,, \varepsilon_-^{~-})\,, \quad \,
\varepsilon^{\,i}_{(2)}=(\varepsilon_+^{~+}\,,-\i\varepsilon_-^{~-})\,, \quad \,
\varepsilon^{\,i}_{(3)}= \, -(\varepsilon_-^{~-}\,,\i\varepsilon_+^{~+})\,, \quad \,
\varepsilon^{\,i}_{(4)}= \,-(\i\varepsilon_-^{~-}\,,\varepsilon_+^{~+})\,,
\ee
where
\be 
\begin{split}
\label{EuclKSTwist}
\varepsilon_+^{~+} & \=\! \sqrt{\frac{\ell}{2}}\;
	{\rme}^{\frac{1}{2}(1-\Omega)t'_E +\frac{\i}{2} (\psi +\phi' )}
	\left(\! \ba{c}\cosh \frac{\rho}{2} \\ -\sinh\frac{\rho}{2}\ea\right)\!\otimes 
	\!\left( \ba{c}\cos \frac{\theta}{2} \\ \sin\frac{\theta}{2}\ea \right)\,, \\
\varepsilon_-^{~-} & \= \!\sqrt{\frac{\ell}{2}}\;
	{\rm e}^{-\frac{1}{2}(1-\Omega)t'_E -\frac{\i}{2} (\psi +\phi' )}
	\left( \!\ba{c}-\sinh \frac{\rho}{2} \\ \cosh\frac{\rho}{2}\ea \right)\!\otimes 
	\! \left( \!\ba{c}-\sin \frac{\theta}{2} \\ \cos\frac{\theta}{2}\ea\right)  \,,
\end{split}
\ee
respect the periodicity~\eqref{Newperiodicity} (they are periodic around the non-conctractible circle and 
anti-periodic around the contractible circle). 

One could also directly solve the new Killing spinor equations~\eqref{TwistKSE}. 
The only non-trivial change compared to the untwisted case is the equation in the~$t'_E$ direction,
\beqa
0&\=&  \bigl( 2\partial_{t'_E} - \omega_{t'_E}^{12}\gamma_{12}- \omega_{t'_E}^{34}\gamma_{34} \bigr) \,\varepsilon_\pm{}^\pm 
-\frac{\i}{2\ell}  E_{t'_E}{}^1 \gamma_{45}\gamma_1 \, \varepsilon_\pm{}^\pm 
-\frac{\i}{\ell} E_{t'_E}{}^5 ( \gamma_{45}\gamma_5 ) \, \varepsilon_\pm{}^\pm \,.
\eeqa
Comparing to the equation \eqref{TwistKSE} that would be written in untwisted  frame, 
The difference with the equation in the untwisted frame is that $2\partial_{t'_E}$ acting on the Killing 
spinor \eqref{EuclKSTwist} gives $\pm(1-\Omega) $ instead of $\pm 1$. 
Also,  the third  and the last terms are new. 
By the projection property along S$^2$ direction of the Killing 
spinor $(1\otimes {\rm e}^{- \i \tau_2 \theta}\tau_3) \ve_\pm{}^\pm = \pm \ve_\pm{}^{\pm}$, 
one can check that the  effect of the third and the last term indeed cancels 
the contribution of $\Omega$ in the time-derivative acting on the Killing spinor.

\vskip 0.2cm
\ndt
\textbf{Supersymmetry algebra:}

The supercharges~$\cQ_a = \delta(\varepsilon^i_{(a)})$,
with the Killing spinors~$\varepsilon^i_{(a)}$,  $a= 1\,,2\,,3\,,4$
defined in~\eqref{4TwistedSMKS}, obey 
\be \label{QQbartwisted} 
\{ \cQ_a\,, \cQ_b \} \=- 2 \i   \delta_{ab}  (L_0 - J^3)\,, \qquad [ L_0 - J^3\,, \cQ_a]\=0\,.
\ee
Consider the following four supercharges~$G^{i \alpha}_\gamma$,
\be \label{n=4 twistedsupercharges2}
G_+^{++} \,\equiv \, \frac{\i \cQ _{1} +\cQ _{2}}{2} \,,  \quad \; G_-^{--} \,\equiv \, \frac{\cQ _{1} + \i \cQ _{2}}{2} \,, \quad \;
G_-^{+-} \, \equiv \, \frac{-  \cQ _{3}+\i \cQ _{4}}{2}\,, \quad \; G_+^{-+}\, \equiv \, \frac{\i  \cQ _{3} -\cQ _{4}}{2} \,,
\ee
where $\gamma$ is the sign of the $L_0$ eigenvalue, $i$ is the doublet index under the outer automorphism 
coming from the~SU(2) R-symmetry of the supergravity, and $\alpha$ is the doublet index under the~SU(2) 
R-symmetry arising from the isometry of the S$^2$. 
They are charged under the bosonic generators~$L_0$ and~$J^3$ as
\be \label{LJGcom} 
 \left[L_0 \,, G^{i \pm}_{\pm} \right] \= \pm \half \,G^{i\pm}_{\pm}\,, \qquad 
 \left[J^3 \,, G^{i \pm}_{\pm} \right] \=  \pm \half \,G^{i \pm}_{\pm}\,,
\ee
so that
\be \label{LJGcomm2}
\bigl[L_0 - J^3 \,, G^{i \pm}_{\pm}\bigr] \=0\,,
\ee
and they obey the anticommutation relations
\be
 \label{Giag_anticom}
\bigl\{G^{+\pm}_{\pm}\,,G^{- \mp}_{\mp} \bigr\} \= \pm   \left( L_0 - J^3\right)\,, \qquad
\bigl\{ G^{+\pm}_\pm \,, G^{-\pm}_\pm \bigr\}\=0\,. 
\ee

The above algebra~\eqref{LJGcomm2}, \eqref{Giag_anticom} forms a subalgebra of the 
global part of $\cN=4$ superconformal algebra in the NS sector given in Section~\ref{LorentzianKSandAlgebra}. 
Note that the subalgebra can also be thought of as the spectral flow\footnote{The spectral flow is taken 
on the charges with algebra in~\cite{Guo:2022ifr} as~$\cL_n \mapsto 
\cL_n +2\eta J^3_n+\eta^2 \frac{c}{6}\delta_{n,0}$, ${J^3_m \mapsto 
J^3_m+\eta \frac{c}{6}\delta_{n,0}}$, {$J^\pm_m \mapsto J^\pm_{m\pm 2\eta}$}, $\cG^{\pm}_{\dot A, r} 
\mapsto \cG^{\pm}_{\dot A, r\pm\eta}$.}, with parameter~$\eta = 1/2$, to the following Ramond sector 
zero-modes as $L_0 - J^3+c/24 \mapsto {\cL}_0^R$, $G^{\pm \mp}_\mp \mapsto 
\mp \cG^\mp_{\mp, 0}$, $G^{\pm \pm}_\pm \mapsto \pm \cG^\pm_{\mp, 0}$.

\vskip 0.2cm

\subsection{The trace interpretation and the semiclassical limit}\label{sec:Semiclassics}
We now interpret the functional integral on  $\Hthrees$  as a trace on the Hilbert space of the boundary CFT. First, a small change of  {\bf conventions} in this section: we consider $(0,4)$ SCFT$_2$ on the boundary, where supersymmetry algebra acts on the right-moving sector, whose generators  we denote with a bar e.g.~$\overline{L}_0$ and $\overline{J}^{\,3}$. This means we have to flip the convention (regarding the bars) of the previous subsection so that we have the following representation  in the twisted coordinates,
\be
{L}_0 \= \i \frac{1}{2} (\i \partial_{t'_E} - \partial_\psi  + \Omega \,\partial_{\phi'}) \,, \qquad \overline{L}_0 \= \i \frac{1}{2} (\i \partial_{t'_E}+ \partial_\psi  + \Omega \,\partial_{\phi'})\,, \qquad 
{\overline J}^{\,3} \= \i \partial_{\phi'}\,,
\ee
with~$\Omega=1+ \i \t_1/\t_2$.
Now consider the gravitational functional integral~$Z$ corresponding to the partition function 
on the twisted torus~\eqref{Newperiodicity}, \eqref{TwistedFrame}. 
Bosonic fields are periodic around both the cycles of the torus, while fermionic fields are 
periodic around the~$t'_E$-circle (which has periodicity~$2 \pi \t_2$) and are 
anti-periodic around the contractible~$\psi$ circle (which has periodicity~$2 \pi$). 
In addition, we have the chemical potential~$2 \pi \i \t_1$ for the angular momentum~$\partial_\psi$ 
around the~AdS$_3$, and the chemical potentials~$ \mu^I$ coupling to~$U(1)$ current(s)~$q_I$.   
By the usual interpretation of the functional integral, we have
\begin{eqnarray}\label{tracefromPI}
\rme^{C(\t,\mu)} \, Z(\t,\mu) & \= & \Tr_\text{NS} \, (-1)^F \, \exp \bigl(  2\pi  \t_2 \, \partial_{t'_E}  
+ 2\pi  \t_1 \, \partial_{\psi} +  \mu^I q_I \bigr) \,, \nn \\
& \= &\Tr_\text{NS} \, (-1)^F \, \exp \bigl( -2\pi  \t_2 \, (L_0+\overline{L}_0- \Omega \overline{J}^{\,3})  
+ 2\pi \i \t_1 \, ({L}_0- \overline{L}_0) + \mu^I q_I  \bigr) \,, \nn \\
& \= & \Tr_\text{NS} \, (-1)^F \, {q}^{{L}_0} \, \overline{q}^{\overline{L}_0-\overline{J}^{\,3}} \,  \rme^{ \mu^I q_I}\,,
\end{eqnarray}
with~$q=\rme^{2\pi \i \t}$, $\overline{q} = \rme^{-2\pi \i \overline{\t}}$, and~$\t = \t_1 + \i \t_2$. 
The Casimir-energy-type term~$C(\t,\mu)$ arises while relating the functional integral to the Hamiltonian 
trace~\cite{Kraus:2006wn}. 

We  recognize the right-hand side of~\eqref{tracefromPI} as the elliptic genus. Indeed, 
from the anticommutator~\eqref{Giag_anticom} we see that the above trace can be written as 
\be \label{traceEllGenfromPI}
\rme^{C(\t,\mu)} \, Z(\t,\mu)  \= \Tr_\text{NS} \, (-1)^F \, {q}^{{L}_0} \, \overline{q}^{ \i \cQ^2 } \,  \rme^{\mu^I q_I} \,,
\ee
where we have chosen one supercharge~$\cQ = \frac1{\sqrt{2}} \cQ_1 =\frac{1}{\sqrt{2}} (G_-^{--} - \i G_+^{++})$ as in \eqref{QQbartwisted}  and~\eqref{n=4 twistedsupercharges2}.  
The pairing of all non-BPS modes with respect to the supercharge~$\cQ$ enforces that 
the elliptic genus is a holomorphic function of~$\t$. 
Note that there is no regularization scheme for the functional integral in which the pre-factor~$C(\t,\mu)$
respects modular invariance and holomorphy. In particular, if one chooses the pre-factor~$C(\t,\mu)$ to 
respect modular invariance (and the gauge invariance associated to~$\mu$), it suffers from a holomorphic 
anomaly and  cannot be purely holomorphic in~$\tau$~\cite{Closset:2019ucb}.


\bigskip

\ndt {\bf On-shell action}

\smallskip

Now that we have set up the twisted torus background, we can evaluate the functional integral 
as explained in Section~\ref{adscft} for the 3d untwisted theory. We have 
\be \label{PI}
Z_\text{AdS}(\tau,\,\mu) \= \int D \phi_\text{grav} \; \rme^{-S_{\text{ren}}(\tau,\,\mu)} \,,
\qquad 
S_{\text{ren}} \,  \equiv \,  S_{\text{bulk}}+S_{\text{bdry}} \,.
\ee
as an integral over all field configurations~$\phi_\text{grav}$ of the 5d supergravity theory, with 
the renormalized action~$S_{\text{ren}}$ of the gravitational theory.

The bulk supergravity action~\eqref{bulk_action} evaluated on the twisted torus~\eqref{TwistedFrame} is 
\be \label{5dbulkeval}
\begin{split}
S_{\text{bulk}}(\tau_2,p,\mu) & \=\frac{1}{8\pi^2}\int_{0}^{\rho_0}d\rho
 \int_{0}^{2\pi}d\psi\int_{0}^{\pi}d\theta\int_0^{2\pi}d\phi'\int_0^{2\pi \tau_2}dt_E'   \,L_{\text{bulk}}\,,  \\
&\= -\frac{\pi \tau_2}{3} p^3 + \text{O}(e^{2\rho_0}) \,.
\end{split}
\ee
The second term on the right-hand side denotes 
terms in the bulk action that diverge when the radial cutoff~$\rho_0 \to \infty$, 
and is absorbed by standard boundary terms. 

The boundary terms in the action of the gauge fields behave essentially in the same way as in the untwisted theory, but 
with slightly different details. 
In the cylinder frame~\eqref{Emetric}, the gauge fields~$W_{z, \bar{z}}$ on the~AdS$_3$ factor 
have the boundary conditions~\eqref{CSbc}, while the components~$W_{\theta, \phi}$ on the~S$^2$ are fixed at 
the boundary.
Twisting these boundary conditions using~\eqref{twistedcoord} gives the boundary conditions 
for the gauge fields on the twisted torus: 
\be \label{CSbctw}
\delta {W}^{I \, (0)}_{z'}\=0\,, \qquad {W}^{I\,(0)}_{\bar{z}'} \; \text{not fixed,} \qquad \delta W^{I \, (0)}_{\theta, \phi'} \= 0 \,,
\ee
where the~$(0)$ indicates the boundary values in the large-$\rho$ expansion as in~\eqref{FGexpQ}.
The Chern-Simons boundary action consistent with these boundary conditions is: 
\beqa  \label{CSbdr}
\begin{split}
S_{\text{CS}}^{\text{bdry}}&= -c_{IJK}\frac{\i p^I}{48\pi^2}\int_{\partial \cM} dz' d\bar{z}'d\theta d\phi'\, 
\sin \theta \Big[(W_{z'}^J-\frac{1}{2}\Omega W^J_{\phi'})W^K_{\bar{z}'}\Big]_{\text{bdry}}\,.
\end{split}
\eeqa
which on the twisted torus~\eqref{TwistedFrame} evaluates to\footnote{Evaluating actions of this type is 
more conviently done by transforming back from the~$(z',\,\bar{z}')$ to the~$(\psi,\,t_E')$ coordinates 
where the ranges are as in~\eqref{5dbulkeval}.}
\be
S_{CS}^{\text{bdry}} \= -\frac{2 \pi \tau_2}{3}c_{IJK}\,\mu^I \mu^J p^K\,.
\ee

The boundary terms in the gravitational sector are more subtle. 
Recall that in the standard treatment of AdS$_3$ gravity,~see e.g.~\cite{Kraus:2006wn}, 
the variation of the boundary metric is fixed, but not that of its normal derivative.
As is well known, one requires the addition of a Gibbons-Hawking boundary term, in order for the 
variational principle to be well-defined. In addition, one requires additional local counterterms on the 
boundary to cancel the divergences arising from the bulk action as in~\eqref{5dbulkeval} as well as 
from the Gibbons-Hawking term.
These considerations are systematically summarized in the procedure of holographic 
renormalization~\cite{Bianchi:2001kw}.
As it turns out, in the localization of the path integral for the elliptic genus, 
we need to impose slightly different boundary conditions on the metric compared to the standard ones, 
and correspondingly we have a different structure of boundary terms. 
However, these differences are only relevant when the metric goes off-shell, and do not change the on-shell 
background that we have discussed so far. 
We postpone the details to a follow-up publication.
Thus the value of the full renormalized action on the twisted background~\eqref{TwistedFrame} is
\be \label{5dactioneval}
 S_{\text{ren}}  \=  -\pi k \tau_2-\pi \tau_2 \,k_{IJ} \, \mu^I \mu^J\,.
\ee

Note that the twisting procedure only affects global properties and does not change the Newton's constant.
Therefore the central charge continues to be~$c=6k=2 p^3$ as in~\eqref{cgravval}.  
Similarly, the level~$k_{IJ}$ of the boundary current algebra also does not change. To see this, note that 
the relation between the twisted and cylinder-frame fields is:
\be
W_{z'} \= {W}_{z}+\frac{1}{2} \Omega \,W^I_{\phi}(\theta)\,,\qquad W_{\bar{z}'} 
\= {W}_{\bar{z}}-\frac{1}{2} \Omega \, W^I_{\phi}(\theta)\,,\qquad W_{\phi'}=W_{\phi}(\theta)\,,
\ee
where~$W_{z,\bar{z}}$ only depends on the AdS$_3$ coordinates~$(\rho, z, \bar{z})=(\rho, z', \bar{z}')$ while~$W^I_{\phi}=-p^I \cos \theta$.
Substituting this into~\eqref{CSbdr} gives:
\be \label{CSbdrReduced}
S_{\text{CS}}^{\text{bdry}} \= -c_{IJK}\frac{\i p^I}{12\pi}\int_{\partial \cM} dz' d\bar{z}' \Big[W_{z}^J W^K_{\bar{z}}\Big]_{\text{bdry}}\,,
\ee
which is the same as the 3d Chern-Simons boundary term~\eqref{CSbdry}, 
since the integration ranges of~$(z',\,\bar{z}')$ are the same as for~$(z,\bar{z})$.
This shows that~$ k_{IJ} \= \frac{2}{3}c_{IJK}\,p^K$ as in the untwisted theory~\eqref{kIJ}.

\section{The Euclidean 4d/5d lift} \label{section:Euc4d5dlift}

In this section we present a formalism to obtain off-shell localization solutions in 5d supergravity 
by lifting the localization manifold around Euclidean~AdS$_2 \,\times \,$S$^2$.
In particular, this allows us to obtain localization solutions around the supersymmetric  
twisted torus~$\Hthrees$ background presented in~\eqref{TwistedFrame}.

We briefly recall the first step of the localization problem that the formalism addresses.
We define the localization supercharge~$\cQ = \frac{1}{\sqrt{2}} \cQ_1 = \frac{1}{\sqrt{2}}\delta(\varepsilon^i_{(1)})$ 
where the Killing spinor~$\ve^i_{(1)}$ is given 
in~\eqref{4TwistedSMKS}. (Equivalently,~$\cQ = \frac{1}{\sqrt{2}} (G_-^{--} - \i G_+^{++})$ in terms 
of the super-Virasoro generators.) It obeys the algebra 
\be \label{defCQ}
\cQ^2 \= - \i  (\overline{L}_0 - \overline{J}^{\,3}) \,, \qquad [ \overline{L}_0 - \overline{J}^{\,3}\,, \cQ]\=0\,.
\ee
We would like to study the space of solutions to the BPS equations 
given by setting the supersymmetry variations generated by~$\cQ$ of all the fermions~\eqref{KSequation1} to zero.

The BPS equations form a system of matrix-valued partial differential equations in terms of the bosonic fields of the theory.
One systematic approach to solve them, assuming no fermionic backgrounds, 
begins by forming various Killing spinor bilinears~\cite{Tod:1983pm, Gauntlett:2002nw}.
The BPS equations may then be expressed as a set of coupled first-order equations for these tensor fields, 
which describe the bosonic background of the solution. This approach was used in~\cite{Gupta:2012cy, Gupta:2019xac} 
to solve the off-shell problem in the~AdS$_2\, \times\,$S$^{2}$ (and~S$^3$) background.   
The general solutions to the resulting equations are, however, typically difficult to obtain, and
we do not solve this problem of general classification in this paper.
Instead, we leverage what is already known about the localization solutions in 4d supergravity around the 
Euclidean AdS$_2 \times$S$^2$ background~\cite{Dabholkar:2010uh, Gupta:2012cy, Gupta:2015gga}, 
by lifting them to five dimensions.
This involves the KK lift of AdS$_2 \times$ S$^2$ to AdS$_3 \times$ S$^2$, which we describe in Section~\ref{section:KK}.
Note, however, that while the 4d localization manifold has been determined completely, there may be additional solutions 
in 5d that \emph{do} depend on the KK direction, and that will therefore not emerge from the lift.
We postpone the discussion of such solutions to future work.

To lift the 4d localization solutions, we use the idea of the 4d/5d off-shell connection of~\cite{Banerjee:2011ts}. 
However, as mentioned in the introduction, implementing this idea  is not straightforward for the following reasons.
Firstly, while the formalism in~\cite{Banerjee:2011ts} was developed for Lorentzian supergravities, 
our 4d/5d connection needs to be adapted to accommodate the Euclidean supergravities in both four and five dimensions.  
A subtlety here, as we will shortly see, is that the 4d Euclidean theory has a redundancy in the choice of reality conditions 
and correspondingly a redundancy of AdS$_2\times$S$^2$ backgrounds, which has no counterpart in the 5d theory. 
Secondly, recall that the 4d/5d lift produces a five-dimensional background in the Kaluza-Klein anstaz and so, in order to 
reach the five-dimensional theory on the supersymmetric twisted torus  $\Hthrees$ from the four-dimensional theory 
on AdS$_2\times$S$^2$, we require a mapping of the twisted torus~\eqref{TwistedFrame} into the Kaluza-Klein frame 
of AdS$_3\times$S$^2$.
In Section~\ref{section:KK} we present the mapping from the Kaluza-Klein frame to the cylinder frame. 
The twisted frame can then easily be mapped to the cylinder frame~\eqref{Emetric} 
by the local coordinate transformation~\eqref{twistedcoord}. 
In Section~\ref{connection} we review the  4d Euclidean supergravity and the \hbox{AdS$_2\times$S$^2$} background. 
In Section~\ref{4d5dconnection}, we present our construction of the Euclidean 4d/5d off-shell lift.
Further, we show that the redundancy of the 4d theory mentioned above can be absorbed into the mapping parameter.
We conclude the section by presenting the steps of lifting the 4d off-shell solutions to the 5d twisted torus.

\subsection{The Kaluza-Klein coordinate frame   \label{section:KK}}

In this subsection we map the cylinder frame to the Kaluza-Klein frame. 
This mapping requires the local coordinate transformations as well as local Lorentz transformations. 
After presenting the general mechanism, we find the specific coordinate and Lorentz transformations, 
and the resulting background configuration and supercharges  
for~AdS$_3 \times $ S$^2$  in the Kaluza-Klein frame.

The general mechanism is as follows. 
Let~$\{\dot{M},\dot{N},\cdots\}$ and~$\{\dot{A},\dot{B},\cdots\}$ be the spacetime and tangent indices,
respectively, in this Kaluza-Klein frame. 
The vielbein in the Euclidean cylinder frame~$E_{M}{}^{A}$ maps to the vielbein in the KK frame $\dot{ E}_{\dot{N}}{}^{\dot{A}}$ 
under a diffeomorphism together with some local rotation~$L_{{A}}{}^{\dot{A}}$ which acts on the frame as~\cite{Scherk:1979zr} 
\be \label{vielbeintransf}
E_{M}{}^{A}(x)\= \frac{\partial \dot{x}^{\dot{N}}}{\partial x^{{M}} }\dot{ E}_{\dot{N}}{}^{\dot{A}}(\dot{x}) L^{-1}_{~\dot{A}}{}^{{A}}\,.
\ee
Correspondingly, the spin connection transforms as
\be\label{omegainKK}
\omega_{{M} {A}}{}^{{B}} \= \frac{\partial \dot{x}^{\dot{N}}}{\partial x^{{M}}} 
\left(L_{{A}}{}^{\dot{A}}\dot{\omega}_{\dot{N} \dot{A}}{}^{\dot{B}} L^{-1}_{~\dot{B}}{}^{{B}} 
+(\partial_{N} L_{{A}}{}^{\dot{A}})\, L^{-1}_{~\dot{A}}{}^{{B}}\right)\,.
\ee
Likewise, the remaining non-trivial background fields and the Killing spinors are mapped into 
the KK frame using the same diffeomorphism and local rotation $L_{{A}}{}^{\dot{A}}$, and a corresponding 
spinor rotation $\cL$, as
\be  \label{VKKtoGlobal}
T_{{A}{B}}\= L_{{A}}{}^{\dot{A}}L_{{B}}{}^{\dot{B}}\dot{T}_{\dot{A}\dot{B}}\,, \qquad 
F_{{A}{B}}\= L_{{A}}{}^{\dot{A}}L_{{B}}{}^{\dot{B}}\dot{F}_{\dot{A}\dot{B}}\,, \qquad 
\epsilon^i \=\cL \,\dot{\varepsilon}^j  \,,
\ee
where  $L_{{A}}{}^{\dot{A}}$ and $\cL$ are related 
such that the gamma matrix is preserved:
\be\label{LorentzSpin}
L_{A}{}^{\dot{B}}\,\cL \, \gamma_{\dot{B}} \, \cL^{-1} \= \gamma_A\,.
\ee

The diffeomorphism and local rotation in~\eqref{vielbeintransf} should be chosen such that the 
vielbein in KK frame~$\dot{ E}_{\dot{N}}{}^{\dot{A}}$ has the following reduction ansatz. Decomposing 
the KK frame coordinate as~$x^{\dot{M}}= \{x^\mu,\, x^5\}$ and $x^{\dot{A}}= \{x^a\,, 5\}$, the reduction 
ansatz of the vielbein is
\be \label{KK_ansatz_vielbein}
{\dot{E}}_{\dot{M}}{}^{\dot{A}}\=\Biggl( \, \begin{matrix}e_\mu{}^a & B_\mu \phi^{-1}\\ 0&\phi^{-1} \end{matrix}\, 
\Biggr)\,, \qquad 
\dot{E}_{\dot{A}}{}^{\dot{M}} \= \Biggl( \, \begin{matrix}e_a{}^\mu & - e_a{}^\mu B_\mu \\ 0&\phi \end{matrix}\, \Biggr)\,,
\ee
where all the fields in the KK frame are independent of the compactified~$x^5$ coordinate. 
Note that the KK ansatz \eqref{KK_ansatz_vielbein} breaks the 5d diffeomorphisms to 4d diffeomorphisms 
and a~$U(1)_{\text{gauge}}$, and breaks the 5d local rotation symmetry $O(5)$ to $O(4) \times \mathbb{Z}_2$. 
Using the $\mathbb{Z}_2$ we can fix the $\phi$ to have a fixed sign, say, positive.
The vielbein \eqref{KK_ansatz_vielbein} is equivalent to 
the following metric in the KK frame (with~$x^5\,\sim\,x^5+ 2 \pi$), 
\be \label{KK_ansatz}
\dot{G}_{\dot{M}\dot{N}}\,dx^{\dot{M}} dx^{\dot{N}} \= g_{\mu \nu}\,dx^\mu dx^\nu +\phi^{-2}(dx^5+B_\mu \, dx^\mu)^2\,.
\ee
We see from the \eqref{KK_ansatz_vielbein} and \eqref{KK_ansatz} that the five-dimensional vielbein $\dot{ E}_{\dot{N}}{}^{\dot{A}}$ 
or  metric~$\dot{G}_{\dot{M}\dot{N}}$ are related to the four-dimensional veilbein~$e_\mu{}^a$  or metric~$g_{\mu\nu}$, 
a gauge field $B_\mu$ and a scalar~$\phi$.
The reduction ansatz leads to the following reduction of the spin connection as
\be\label{omegareduction}
\dot{\omega}_{\dot{A} }{}^{bc}\= \Biggl( \, \begin{matrix}\omega_a{}^{bc} \\ \frac{1}{2}\phi^{-1} F(B)^{bc} \end{matrix}\, 
\Biggr)\, \qquad \quad 
\dot{\omega}_{\dot{A} }{}^{b5}\= \Biggl( \, \begin{matrix} -\frac{1}{2} \phi^{-1}F(B)_a{}^{b} \\ 
- \phi^{-1} D^{b}\phi \end{matrix}\, \Biggr)\,.
\ee
Here we see that the gauge field $B_\mu$ appears through its field strength  $F(B)_{ab}$ in four dimensions.

Now we find the coordinate transformations and the local rotation in~\eqref{vielbeintransf}  
that map the cylinder frame background in~\eqref{Emetric} to fit into the KK frame ansatz~\eqref{KK_ansatz_vielbein} 
and~\eqref{KK_ansatz}. 
The cylinder frame coordinates $x^M$ and and the KK frame coordinates~$\dot{x}^{\dot{M}}$ 
\be\label{GlobaltoKK_coord}
x^M \=( \rho \,, \psi\,, \theta\,,\phi\,, t_E)\,,~~~\dot{x}^{\dot{M}} \= (\eta\,, \chi\,, \theta\,,\phi\,, x^5)\,,
\ee
are related as 
\be \label{KKcoords}
(\, \rho, \psi, t_{E} ) \= 
\Bigl(\, \frac{\eta}{2} , \, \chi +\i \frac{x^5}{2}, \, \frac{x^5}{2} \,\Bigr) 
\quad \Leftrightarrow \quad 
 (\eta \,, \chi\,, x^5) \= (2\rho \,, \psi-\i t_E \,, 2  t_E )\,,
\ee
with the coordinates~$(\theta, \phi)$ remaining the same. 
Note that the global conditions on the periodicities are not respected by this map (e.g.~$x^5$ is compact whereas $t_E$ is not).
The corresponding local rotation matrix~$L_{A}{}^{\dot{A}}$  
is given as a rotation in the~$2-5 $ plane (along $\hat{\psi}$ and $ \hat{t}_E$ direction)  
with angle~$\omega = -\i \eta/2$:
\be\label{ToKK_Loreantz}
L_{{1}}{}^{\dot{1}}= L_{{3}}{}^{\dot{3}}= L_{{4}}{}^{\dot{4}}\=1\,,\qquad 
L_{{2}}{}^{\dot{2}}=  L_{{5}}{}^{\dot{5}}=\cosh\frac{\eta}{2}\,,\qquad 
L_{{2}}{}^{\dot{5}}=  -L_{{5}}{}^{\dot{2}}\=\i \sinh\frac{\eta}{2}\,.
\ee
In the exponential form,  we have~$L_{A}{}^{\dot{A}}= ({\rm e}^{\Omega})_{A}{}^{\dot{A}}$ , 
where the $2-5$ component of the matrix in the exponent is~$\Omega_{25} =-\Omega_{52}= -\omega = \i\eta/2$. 
\footnote{ The  rotation with angle $\omega$ is $\exp\biggl( \omega \Bigl(\begin{smallmatrix} 0&-1\\1&0\end{smallmatrix} \Bigr)\biggr) = 
\Bigl(\begin{smallmatrix} \cos\omega&-\sin\omega\\\sin\omega&\cos\omega \end{smallmatrix} \Bigr)$. 
Here, we take $\omega = -\i \eta/2$ for the  rotation in $2$--$5$ plane.
Note that the angle is imaginary, because the coordinate~$x^5$ is Euclidean time.} 
By the relation \eqref{LorentzSpin}, the corresponding  spinor rotation is
\be \label{GlobaltoKKSpin}
\begin{split}
\cL & \= \exp \left(\frac{1}{4}\Omega_{AB}\gamma^{AB}\right) = \exp \left( \frac{\i}{4} \eta \,\gamma_{25}\right)
=\Biggl( \, \begin{matrix} \cosh \frac{\eta}{4} & -\sinh\frac{\eta}{4}\\ 
-\sinh\frac{\eta}{4}&\cosh \frac{\eta}{4} \end{matrix}\, \Biggr)\otimes \mathbb{I}_2\,.
\end{split}
\ee
We note that although the spin connection in the cylinder frame has zero component 
for~$\omega_{\hat{\rho}}{}^{\hat{\psi}\hat{t}_E}$, as can be seen in~\eqref{spinconnection5dLor}, 
the corresponding spin connection of KK frame mapped by \eqref{omegainKK}, 
$\dot{\omega}_{\hat{\eta}}{}^{\hat{\psi}\hat{t}_E} (= \dot{\omega}_{\dot{1}}{}^{\dot{2}\dot{5}})$, 
is non-zero due to the contribution of the  Lorentz transformation matrix in the second term 
of~\eqref{omegainKK}. According to~\eqref{omegareduction}, this non-zero component gives the 
non-zero value of the electric flux along AdS$_2$ .
This explains why there is electric flux on AdS$_2$ even though the AdS$_3$ does not have any electric flux.

\medskip

We now summarize the on-shell supersymmetric field configuration in the KK coordinates. 
By using the transformations \eqref{KKcoords}  \eqref{ToKK_Loreantz} \eqref{GlobaltoKKSpin} on 
the background Weyl multiplet as in \eqref{Emetric} and matter multiplets  as in \eqref{vector_field_content} 
\eqref{hyper_field_content}, we obtain the following  configuration:
\beqa\label{backgroundE5d}
    d{{s}_5}^2 && \= \ell^2\,(d\eta^2+\sinh^2{\eta}\,d\chi^2+d\theta^2+\sin^2{\theta}\,d\phi^2)
    +\ell^2\,(dx^5+\i  (\cosh{\eta}-1)\,d\chi)^2\,, \nonumber\\
 \dot{T}_{34}&& \=- \frac{1}{4\ell}\,,
 \\
\dot{\sigma}^I&& \=-\frac{p^I}{\R}\,,~~~~~~~\dot{F}^I_{{\theta}{\phi}} \= {p^I} \sin \theta\,,~~~~~~ ~~~\dot{W}^I_{x^5} \= 
\mu^I\,  \,,
\nonumber \\
\dot{A}_1{}^1 && \= \dot{A}_2{}^2 \=  \sqrt{\frac{p^3}{3\ell^3}}
\nonumber\,.
\eeqa
We note that the background geometry has an S$^1$ fibration over the four-dimensional base, 
which is Euclidean~AdS$_2 \times$S$^2$. 
The angular coordinate $\chi$ of Euclidean~AdS$_2$ has periodicity~$2\pi$.
By comparing the metric with the KK ansatz \eqref{KK_ansatz}, 
we identify the following values for the KK one-form and scalar:
\be \label{phi_and_B}
     B\=\i (\cosh{\eta}-1)\,d\chi\quad, \quad  \phi\=\ell^{-1}\,.
\ee

The background configuration given in~\eqref{backgroundE5d} has well-defined supersymmetry. 
To see that, we look for the Killing spinors. By the Eulidean continuation of the Lorentzian Killing 
spinors~\eqref{8SMKS} followed by the coordinate transformation~\eqref{KKcoords} and the 
Lorentz transformation~\eqref{GlobaltoKKSpin} we obtain 
\be\ba{ll}\label{4SMKSinKK}
\dot{\varepsilon}^{\,i}_{(1)}\=(-\i \dot\varepsilon_+^{~+}\,, \dot\varepsilon_-^{~-})\,, ~~~~~~&
\dot\varepsilon^{\,i}_{(2)}\=(\dot\varepsilon_+^{~+}\,,-\i\dot\varepsilon_-^{~-})\,,\\
\dot\varepsilon^{\,i}_{(3)}\=(-\dot\varepsilon_-^{~-}\,,-\i\dot\varepsilon_+^{~+})\,, &
\dot\varepsilon^{\,i}_{(4)}=(-\i\dot\varepsilon_-^{~-}\,,-\dot\varepsilon_+^{~+})\,,\\
\dot{\tilde{\varepsilon}}^{\,i}_{(1)}\=(\dot\varepsilon_+^{~-}\,,\i\dot\varepsilon_-^{~+})\,,&
\dot{\tilde\varepsilon}^{\,i}_{(2)}\=(\i\dot\varepsilon_+^{~-}\,,\dot\varepsilon_-^{~+})\,,\\
\dot{\tilde\varepsilon}^{\,i}_{(3)}\=(-\i\dot\varepsilon_-^{~+}\,,\dot\varepsilon_+^{~-})\,,&
\dot{\tilde\varepsilon}^{\,i}_{(4)}\=(\dot\varepsilon_-^{~+}\,,-\i\dot\varepsilon_+^{~-})\,.
\ea\ee
where the spinors $\dot\varepsilon_\pm^{~\pm}$ and $\dot\varepsilon_{\pm}^{~\mp}$ are 
\be \nonumber
\dot\varepsilon_+^{~+}\=\! \sqrt{\frac{\ell}{2}}\,{\rm e}^{\frac{\i}{2} (\chi +\phi)}
\Biggl(\! \ba{c}\cosh \frac{\eta}{2} \\ -\sinh\frac{\eta}{2}\ea\Biggr)\!\otimes \!
\Biggl( \ba{c}\cos \frac{\theta}{2} \\ \sin\frac{\theta}{2}\ea\Biggr)\,, \quad
\dot\varepsilon_+^{~-}\=\! \sqrt{\frac{\ell}{2}}\,{\rm e}^{\frac{\i}{2} (\chi -\phi)}
\Biggl( \!\ba{c}\cosh \frac{\eta}{2} \\ -\sinh\frac{\eta}{2}\ea\Biggr)\!\otimes \!
\Biggl( \!\ba{c}-\sin \frac{\theta}{2} \\ \cos\frac{\theta}{2}\ea\Biggr) \,,
\ee
\be  \label{4KSKKframe} 
\dot\varepsilon_-^{~+}\=\!\sqrt{\frac{\ell}{2}}\,{\rm e}^{-\frac{\i}{2} (\chi -\phi)}
\Biggl(\! \ba{c}-\sinh \frac{\eta}{2} \\ \cosh\frac{\eta}{2}\ea\Biggr)\!\otimes \!
\Biggl( \ba{c}\cos \frac{\theta}{2} \\ \sin\frac{\theta}{2}\ea\Biggr) \,, \quad 
\dot\varepsilon_-^{~-}\= \!\sqrt{\frac{\ell}{2}}\,{\rm e}^{-\frac{\i}{2} (\chi +\phi)}
\Biggl( \!\ba{c}-\sinh \frac{\eta}{2} \\ \cosh\frac{\eta}{2}\ea\Biggr)\!\otimes \!
\Biggl( \!\ba{c}-\sin \frac{\theta}{2} \\ \cos\frac{\theta}{2}\ea\Biggr)  \,.
\ee
Note that they are well-defined with respect to the global structure of the geometry~\eqref{KK_ansatz} 
because they do not depend on the~$x^5$ direction (the spinors above are in fact precisely 
the four-dimensional Killing spinors on AdS$_2 \times $ S$^2$, as we spell out in Appendix~\ref{KSonAdS2S2}).
Note also that, as in the twisted torus frame, we cannot impose any reality conditions on the Euclidean spinors. 
This is because although they formally satisfy
$(\ve^i )^\dagger \i \gamma_{5} \= \varepsilon_{ij}(\ve^j)^T C$, 
which is the same symplectic-Majorana condition of the Lorentzian theory~\eqref{SMLor},
this condition is not compatible with the local rotation  of the Euclidean theory.

\subsection{4d Euclidean supergravity and AdS$_2 \times $S$^2$ background \label{connection}}
The Kaluza-Klein formalism described in the previous subsection naturally connects the 5d supergravity on 
the AdS$_3 \times $S$^2$ background in KK coordinates given in~\eqref{backgroundE5d} to the 4d supergravity 
on an AdS$_2 \times $S$^2$ background. In this subsection, we review the 4d Euclidean conformal supergravity 
and the AdS$_2 \times $S$^2$ background in more detail.
In the 4d Euclidean theory, there is a one-parameter redundancy for describing this background that comes from 
the possible choice of reality condition for the fermions. 
\vskip 0.2cm
\ndt
\textbf{4d $\cN=2$ supergravity:} For the 4d~$\mathcal{N}=2$ Euclidean conformal supergravity, we consider 
 the Weyl multiplet, coupled to~$N_\text{v} +1$ vector multiplets and one hypermultiplet. One of the vector 
 multiplets and the single hypermultiplet act as the compensators to consistently gauge-fix the dilatations of the off-shell theory.
The fields of the Weyl multiplet are
\be
\{{e_\mu}^a \, , \,\psi^i_a \,  ,\,  A_\mu^D \, ,\, A_\mu^R \, ,\, {\cV_\mu}^i{}_j \, ,\,  T_{ab}^{\pm} \, ,\, \cD \, , \,\chi_{4d}^i \}\,,
\ee
 corresponding, respectively, to the vielbein, gravitino, dilatations gauge field, the~$SO(1,1)_R$ and~$SU(2)_R$ 
 gauge fields \footnote{The R-symmetry group of the Euclidean theory is~$ SU(2) \times SO(1,1)$ compared 
 to~$SU(2) \times U(1)$ in the Lorentzian case.}, auxiliary self-dual/anti-self-dual two-form, auxiliary scalar, and 
 the auxiliary fermion. As in the five-dimensional case, we fix~$A^{D}_\mu = 0$ using the K-gauge.
The fields of the~$N_\text{v}+1$ vector multiplets are
\be
\{ X^\cI \, ,\, \overline{X}^\cI \, ,\,   A_\mu^\cI\, , \,\lambda^{\cI\,i} \,  ,\, \cY^{\cI\, ij}\}\,, \qquad \cI \= 0 \,  ,\, \cdots ,\,N_\text{v} \,,
\ee
corresponding to the complex scalar and its conjugate, the~$U(1)$ gauge field, the gaugino, and the auxiliary~$SU(2)$ triplet.
Finally, the hypermultiplet consists of scalars and fermions, 
\be
\{ \cA_i{}^\alpha \, , \, \zeta_{4d}^\alpha \}\,.
\ee
The supersymmetry transformations on the spinor fields~$\psi^i_a ,\,\lambda^{\cI \, i},\, \zeta_{4d}^\alpha$ 
are presented in~\eqref{QSKWeyl}, following the conventions of~\cite{Jeon:2018kec}. 

The  4d $\cN=2$ supergravity is governed by the prepotential $F(X)$ which is homogeneous of degree~2. 
Here, we choose the prepotential as \cite{Banerjee:2011ts} 
\be \label{prepotential}
 F(X) \= -\frac{1}{12}\, c_{IJK}\frac{X^I X^J X^K}{X^0}\,,
\ee
(the sum running over~$I=1,\dots \, N_\text{v}$), 
such that the vector multiplet sector of the 4d theory matches that  of the 5d theory described 
in the section~\ref{5dSUGRA}, according to the 4d/5d map that we will present shortly in 
Subsection~\ref{4d5dconnection}.

\vskip 0.2cm
\ndt
\textbf{Reality conditions:}
Note that in the Euclidean theory the fields~$X^\cI$ and~$\overline{X}^\cI$---and, more generally, fields related 
by complex conjugation in the Lorentzian theory (e.g.~$T^+_{ab}$ and~$T^-_{ab}$)---are independent in the Euclidean theory. 
In order to preserve the number of degrees of freedom, we should impose reality conditions in the Euclidean theory.
This may be done by imposing an appropriate reality condition on the spinors and using supersymmetry. 
The minimal spinors in the $\cN=2$ four-dimensional Euclidean theory can be chosen to obey the 
symplectic-Majorana condition. We note that there are actually an infinite number of such consistent conditions 
which, for any spinor~$\psi^i$, are parametrized by a real number $\alpha$ as
\be \label{allreality}
(\psi^i)^\dagger {\rm e}^{\i \alpha \gamma_5}= \epsilon_{ij} (\psi^i)^T C  \,, \qquad \alpha \in \mathbb{R}\,.
\ee
This infinite choice stems from the fact that the chiral and anti-chiral spinors are independent in Euclidean 4d, and the symplectic-Majorana condition for the chiral and anti-chiral spinors can be imposed with relatively different phases. 
Two natural examples are:
\be \label{4drealityKS}
\alpha \= \pi/2: \quad (\psi^i)^\dagger \i \gamma_5 = \epsilon_{ij}(\psi^j)^T C\,, \qquad \quad 
\alpha = 0: \quad (\psi^i)^\dagger \= \epsilon_{ij}(\psi^j)^T C \,.
\ee
A spinor satisfying the general reality condition~\eqref{allreality} (which we denote by~$\psi^i(\alpha)$)
is related to spinors satisfying~\eqref{4drealityKS} 
\be \label{spinoralpharel}
\psi^i(\alpha) \= {\rm e}^{\frac{\i}{2}\gamma_5(\alpha-\frac{\pi}{2})} \psi^i(\pi/2) \=  {\rm e}^{ \frac{\i}{2} \alpha \gamma_5}\psi^i(0)\,.
\ee

Now, if we impose one such condition on all the spinors of the theory (including the Killing spinors), then 
the consistency of the supersymmetry transformations under this condition fixes specific reality conditions on the bosonic fields.
For the two examples above we have, respectively, the following conditions for the relevant bosonic fields:
\be
\begin{split}
\alpha \= \pi/2: \qquad \bigl(T_{ab}^\pm \bigr)^*  &\= T_{ab}^\pm \;, \quad \quad \bigl(X^\cI \bigr)^*   
\=  X^\cI \,, \qquad \bigl(\overline{X}^\cI \bigr)^*  \=  \overline{X}^\cI \,, \qquad \label{4dreality} \\
\alpha \= 0: \qquad \bigl(T_{ab}^\pm \bigr)^*  &\= -T_{ab}^\pm \;, \quad \; \bigl(X^\cI \bigr)^*   
\=  -X^\cI \,, \quad \, \bigl(\overline{X}^\cI \bigr)^*  \=  -\overline{X}^\cI \,. \qquad 
\end{split}
\ee
However, note that imposing either reality condition in~\eqref{4dreality} causes the wrong sign for the kinetic terms 
of the action, making path integral ill-defined. In fact, this is the case for all bosonic reality conditions implied from 
supersymmetry by~\eqref{allreality}. As was discussed in Section~\ref{5dSUGRA}, the resolution is to impose the 
standard reality condition on the bosonic fluctuations, e.g.~$(\delta X^\cI)^\ast = \delta \overline{X}^\cI$, 
so that path integral is well-defined, and to treat the fermion fluctuations $\psi^1$ and $\psi^2$ as being independent.  
For the background, however, the effect of the choice for $\alpha$ still remains: there is a one-parameter family 
of Killing spinors that satisfy the reality condition \eqref{allreality}, and the supersymmetric bosonic background 
has a corresponding dependence on the choice of $\alpha$ as we will shortly see below.

\vskip 0.2cm
\ndt \textbf{4d AdS$_2\times$S$^2$ background:} 
Here we present the Euclidean AdS$_2 \times$ S$^2$ background, including the complete Weyl multiplet and matter multiplets.
This solution can be obtained by Wick rotation of the Lorentzian AdS$_2 \times$ S$^2$ solution,
which carries both electric and magnetic charges~$\left(q_{\cI},\,p^{\cI} \right)$.
The non-trivial fields are:
\be\label{backgroundE}
\begin{split}
{ds_4}^2 &\= g_{\mu \nu}\,dx^\mu dx^\nu  \= \ell^2 \left(d\eta^2+\sinh^2\eta\, d\chi^2+d\theta^2 +\sin^2\theta d\phi^2\right)\,, \\
T^{-}_{{1}{2}} & \=- \i \,\omega\,,~~~~~~{T}^{+}_{{1}{2}} \=-\i \,\overline{\omega}\,,~~~~~~T^{-}_{{3}{4}} \=
\i \,\omega \,,~~~~~~{T}^{+}_{{3}{4}} \=-\i  \,\overline\omega 
\\
 A^{\cI} &\= - \i\, e^{\cI}(\cosh \eta -1)\,d\chi  -p^{\cI} \cos\theta \, d\phi \,,\\
{X^{\cI}}&\=\frac{\omega}{8}(e^{\cI}+\i p^{\cI})\,,~~~~~\overline{X}^{\cI}\=\frac{\overline{\omega}}{8}(e^{\cI}-\i p^{\cI})\,,\\
 {\mathcal{A}_i}^{\alpha} & \= {a_i}^{\alpha}\=\text{constant} \,,
 \end{split}
\ee
By the field equation for the auxiliary scalar~$\cD$, the ${a_i}^{\alpha}$ are constrained to obey:
\be \label{4dhyperconstraint}
\Omega_{\alpha \beta}\,\varepsilon^{ij}{a_i}^{\alpha}{a_j}^{\beta} \= -4\i(F_{\cI}\overline{X}^\cI-\overline{F}_{\cI} X^\cI)\,.
\ee
By the attractor equations \cite{Sahoo:2006rp},  the electric field $e^\cI$ is related to the electric charge $q_\cI$ as 
\be
4\i \left(\overline{\omega}^{-1}\frac{\partial\overline{F}(\overline{X})}{\partial \overline{X}^\cI}  
- \omega^{-1}\frac{\partial {F}(X)}{\partial {X}^\cI}\right) \=q_\cI\,,
\ee
and  the two independent complex parameters $\omega$ and $\bar\omega$ (unlike in the Lorentzian theory, 
they are not complex conjugate to each other) are related to the length scale of the metric~$\ell$  as 
\be \label{ell}
\ell^2 \= \frac{16}{\omega \overline{\omega}}\,,
\ee
which indeed scales consistently with Weyl weight $(-2)$ and $SO(1,1)_R$ weight $0$.
Since the two complex parameters $\omega$ and $\bar\omega$ carry opposite charges under  
the $SO(1,1)_R$ gauge symmetry, we can set their magnitude to be same:
\be \label{omegamagn}
|\omega| \= |\overline{\omega}|  \= 4/\ell\,.
\ee
Note that to match our 5d set-up, we uphold the dilatational symmetry, which is manifested here in the form of 
an arbitrary value for $\ell$ (one may break the symmetry by fixing $\ell$ to $1$ for instance, as in~\cite{Dabholkar:2010uh}). 
The relation~\eqref{ell},~\eqref{omegamagn} indicates that $\omega$ and $\bar\omega$ are now formally 
conjugate to each other so that we can rewrite them using the following parametrization:
\be  \label{omChoices}
\omega(\alpha) \= \frac{4}{\ell}e^{\i \alpha}\,,\qquad \overline\omega(\alpha)  \= \ \frac{4}{\ell}e^{-\i \alpha }\,,\qquad \alpha \in \mathbb{R}\,.
\ee
Unlike in the 4d Lorentzian theory, where the phase~$\alpha$ is fixed by the~$U(1)_R$ gauge symmetry,
in the Euclidean theory it remains as a free parameter. It is, in fact, precisely the parameter that determines 
the choice of reality condition for the spinors as in \eqref{allreality}, i.e.~the background described in~\eqref{backgroundE} 
with generic~$\alpha$ as in~\eqref{omChoices} preserves the supersymmetries generated by Killing spinors obeying 
the reality condition~\eqref{allreality}. For the case of $\alpha = \pi/2$ the 8 pairs of Killing spinors are presented 
in Appendix~\eqref{8SMKSinKK} and the Killing spinors for a generic~$\alpha$ can be read off from~\eqref{spinoralpharel}. 
Note that the Killing spinors in~\eqref{8SMKSinKK} are exactly same Killing spinors as those of the 5d KK frame 
given in~\eqref{4SMKSinKK}.
 
Now, by comparing the 4d background~\eqref{backgroundE} to the 5d KK-frame background~\eqref{backgroundE5d}, 
it is clear that the AdS$_2\times$ S$^2$ metric in the former is the reduction of the  AdS$_3\times$ S$^2$ metric in the 
latter, as mentioned in Section~\ref{section:KK}. However, it is not yet clear how the 4d/5d background values of the 
other fields are related (beyond just the metric), and how off-shell fluctuations are connected.
In the next subsection, we will elucidate these points by describing the full off-shell map between the Euclidean 4d and 5d supergravity.
Using this map, we will explicitly present how the 4d/5d backgrounds are mapped.

\vskip 0.2cm
\ndt

\vskip 0.2cm

\subsection{The off-shell Euclidean 4d/5d lift}  \label{4d5dconnection}

In this subsection, we describe the off-shell connection between the 4d Euclidean and 5d Euclidean theory. 
We present how the AdS$_2\times$ S$^2$ on-shell background in~\eqref{backgroundE} maps to the AdS$_3\times$ S$^2$ 
on-shell background in KK frame \eqref{backgroundE5d}. This involves a choice of the relevant parameters of the 4d 
background, specifically  $(e^0, p^0)$ in~\eqref{backgroundE}, and depending on the choice of parameter~$\omega$ 
and~$\bar\omega$~\eqref{backgroundE}, a proper mapping parameter is determined. 
We then show how to reach the 5d twisted torus background. 
We end the section with the steps to lift off-shell localization solutions to the 5d twisted torus.

To obtain the Euclidean 4d/5d connection, we use the Lorentzian 4d/5d relations 
of~\cite{Banerjee:2011ts} and map the two theories to their consistent Euclidean counterparts. 
Getting the Euclidean 5d theory by the Wick rotation is straightforward, as explained in Section~\ref{5dSUGRA}. 
We follow the conventions of the 4d Euclidean theory  in~\cite{Jeon:2018kec}. Equivalently, one can start 
from the relations between the 5d Lorentzian and 4d Euclidean theories of \cite{deWit:2017cle}, and Wick 
rotate the 5d theory. The map obtained in this approach differs from ours only in the way that the conventions 
of the 4d Euclidean theory of~\cite{deWit:2017cle} differ from those of the 4d Euclidean theory 
of~\cite{Jeon:2018kec}\footnote{The mapping between these conventions is also presented in \cite{Jeon:2018kec}.}.

Under Kaluza-Klein reduction of the 5d conformal supergravity to 4d, the vector multiplets~$I=1,\dots , N_\text{v}$ 
reduce to the corresponding 4d matter vector multiplets~$\mathcal{I}=1,\dots , N_\text{v}$, and the Weyl multiplet 
reduced to the 4d Weyl multiplet and the additional Kaluza-Klein vector multiplet~$\mathcal{I}=0$.

One can expect that the Kaluza-Klein scalar $\phi$ associated with the 
5d metric~\eqref{KK_ansatz_vielbein} falls into the scalar in the 4d Kaluza-Klein vector multiplet. 
However, directly performing this reduction only gives one real scalar degree-of-freedom, while there should 
be two real degree-of-freedom for the scalars of the vector multiplet. Additionally, the 4d $SO(1,1)_R$ symmetry 
factor is not realized in any of the multiplets. To  recover the missing scalar~d.o.f., an additional field~$\varphi$ is 
introduced~\cite{Banerjee:2011ts, deWit:2017cle} to define the two 4d scalars in the KK vector multiplet as
\be \label{X0rel}
X^0= -\frac{\i}{2}{\rm e}^{-\varphi}\phi\,, \qquad \overline{X}^0= \frac{\i}{2}{\rm e}^{ \varphi}\phi\,.
\ee
The field~$\varphi$ transforms locally under SO(1,1)$_R$ as 
\be \label{SO(1,1)}
\varphi \, \rightarrow \, \varphi +  \Lambda^0\,,
\ee
where~$\Lambda^0$ is real.
One can then consistently couple~$\varphi$ to the remaining 4d fields, so that the $SO(1,1)_R$ of the 4d theory is realized.
 
We now present the explicit 4d/5d mappings, up to quadratic order in the fermions, keeping the general~$\varphi$ dependence.
The 4d Weyl multiplet is related to the 5d Weyl multiplet as: 
 \beqa
 e_\mu{}^{{a}} & \= & \dot{E}_\mu{}^{{a}}\,,\\
{\psi^{i}_{{a}}}&\= & e^{-\frac{1}{2}  \varphi \gamma_5} \dot{\Psi}_{{a}}^{~i}\,,\\
  A^R_{{a}} & \= &- 6\i \dot{T}_{{a}{5}} +  {e_{a}}{}^{\mu}\partial_{\mu}\varphi \label{AR} \\
{\cV_a}^i{}_{j} & \= &{\dot{V}_{a\,j}}{}^i \,,\\
T^{4d \pm}_{{a}{b}} & \= &e^{\pm  \varphi}(24\, \dot{T}^{5d}_{{a}{b}} + \i \phi^{-1}\varepsilon_{{a}{b}{c}{d}}F(B)^{{c}{d}})^{\pm}\,,
\label{Tlift}\\
\cD &\=& 4 \dot{D} +\frac{1}{4}\phi^{-1}e^{a\mu} {D}_\mu ( e_a{}^{\nu }{D}_\nu \phi)+\frac{3}{32}\phi^{-2}F(B)^{ab}F(B)_{ab}
\\
&&\quad\quad + \frac{3}{2}\dot{T}_{\dot{A}\dot{B}}\dot{T}^{\dot{A}\dot{B}}+\frac{1}{4}\phi^2 \dot{V}_{x^5 i }{}^j \dot{V}_{x^5 j}{}^i  \,,
\label{Dlift} \nn
\\
\chi_{4d}^i & \= & 8 \dot\chi^i+\frac{1}{48}\gamma^{ab}F(B)_{ab} 
\dot\Psi^i_{x^5}-\frac{3\i}{4}\phi\dot{T}_{ab}\gamma^5 \gamma^{ab}\dot\Psi_{x^5}^i\\
&& \quad \quad+\frac{1}{4}\phi^{-1}\gamma_5 \slashed{\cD}(\phi^2 \dot\Psi_{x^5}^i)-\frac{1}{2} 
\phi^2 {V_{x^5}}_j{}^i\dot\Psi_{x^5}^j-\frac{9}{4}\i\phi \dot{T}_{a5}\gamma^a\dot\Psi_{x^5}^i\,, \nonumber
\eeqa
where~$\varepsilon_{{a}{b}{c}{d}}$ is the four-dimensional Levi-Civita symbol, and the 4d conformally invariant D'Alembertian is $e^{a\mu}D_\mu (e_{a}{}^\nu D_\nu \phi)= (\cD^a\cD_a +\frac{1}{6}R)\phi$, where $R$ is the 4d Ricci scalar and $\cD_a$ is the 4d Lorentz and R-symmetry covariant derivative.
The 4d supersymmetry parameters are given in terms of the 5d supersymmetry parameters and 5d Weyl multiplet fields as
\beqa
\ve^i_{4d} &\=& {\rm e}^{-\frac{1}{2}\varphi \gamma_5} \dot{\varepsilon}^i \,,
\\
\eta_{4d}^i &\=& -\i \gamma_5 {\rm e}^{\frac{1}{2}\varphi \gamma_5}\left(\dot{\eta}^i - 
2 \dot{T}_{{a}{5}}\gamma^{a} \gamma^{5} \dot{\varepsilon}^i +\frac{\i}{8} \phi^{-1}\gamma_{5} (F(B)_{{a}{b}} -
4\i \phi \dot{T}_{{a}{b}} \gamma_{5})   \gamma^{ab}\dot{\varepsilon}^i\right)\,.
\eeqa
Moving on to the vector multiplets, the 4d KK vector multiplet fields in terms of the 5d Weyl multiplet are:
\beqa
X^0 & \= & -\frac{\i}{2}  e^{-\varphi}\phi\,,~~~~\overline{X}^0\= \frac{\i}{2} e^{\varphi}\phi\,, \label{KKscalar}\\
A^{0}_{{a}} &\=&e_{{a}}{}^{\mu}B_\mu\,,  \label{KKvector}\\
\lambda^{0\,i} & \= &e^{-\frac{1}{2}  \varphi \gamma_5} \dot{\Psi}_{{5}}^{~i} \phi\,,
\\
\cY^{0\,i}{}_j & = & \phi\, {\dot{V}_{5\,j}}{}^i \label{KKY} \,,
\eeqa
and the 4d matter vector multiplet fields in terms of the 5d vector multiplet fields are:
\beqa
X^I & \= & \half e^{- \varphi}(\sigma^I +\i \dot{W}^I_{{5}})\,,~~~\overline{X}^I \=\half e^{ \varphi}(\sigma^I -\i \dot{W}^I_{{5}})\,, \label{XIlift} \\
A^{I}_{{a}} & \= & \dot{W}^{I}_{{a}}\,,\\
\lambda^{I\,i} & \= & e^{-\frac{1}{2} \varphi \gamma_5} \left(\dot{\Omega}^{I\,i} - \dot{W}^I_{{5}} \dot{\Psi}_{{5}}^{~i}\right)\,,\\
\cY^{I\,i}{}_j & \= & -2 \left(Y^{I\,i}{}_{j} +\half \dot{W}^I_{{5}}{\dot{V}_{5\,j}}{}^i \right)\,.
\eeqa
Finally, the 4d hypermultiplet in terms of the 5d hypermultiplet is
\beqa
\cA_i{}^\alpha \= \phi^{-1/2} \dot{A}_i{}^\alpha \label{KKhyper}\,.
\eeqa
Using the above maps, the 4d supersymmetry transformation is obtained from the 5d supersymmetry 
transformation together with a 5d local rotation,
\be
\delta^{4d} \= \delta^{5d}+ \delta_M(\ve)\,, \qquad \ve_{5a } \= -\ve_{a5}\=  \overline{\dot{\varepsilon}}_i \gamma_a \Psi^i_{5}\,,
\ee
where the rotation parameter $\ve_{AB}$ is chosen to fix the gauge $\dot{E}_{x^5}{}^a = \dot{E}_{5}{}^{\mu}=0$.  
We also need the supersymmetry transformation rule of $\varphi$, 
\be
\delta^{5d} \varphi \=  \overline{\dot{\varepsilon}}_i \dot\Psi^i_{5}\,.
\ee

For the purpose of lifting the 4d configuration to 5d, we use the inverse map, namely the 5d fields in terms of the 4d fields. 
The 5d Weyl multiplet fields are given in terms of the 4d Weyl multiplet and 4d KK multiplet as:
 \beqa
  \dot{E}_\mu{}^{{a}} & \=& {e}_\mu{}^{{a}}\,\,,\, \, \dot{E}_\mu{}^{{5}} \= \phi^{-1}B_\mu\,\,,\,\,\, \dot{E}_{x^5}{}^{{5}} \=\phi^{-1} \,,\label{E54}\\
{\dot{\Psi}^{i}_{{a}}}&\=& e^{ \frac{1}{2}\varphi \gamma_5} \psi_{{a}}^{i}\, \, , \,\,\, \dot{\Psi}_{{5}}^{~i}  
\= \phi^{-1} e^{\frac{1}{2}  \varphi \gamma_5} \lambda^{0\,i} \,, \label{gravitino54}\\
\dot{T}_{{a}{b}} & \=& \frac{1}{24}\left(e^{- \varphi}{T^{+}_{{a}{b}}} +e^{ \varphi}T^{-}_{{a}{b}} -
\i \phi^{-1}\varepsilon_{{a}{b}{c}{d}}\,F(B)^{{c}{d}}\right)\,, \label{Tab54}\\ 
\dot{T}_{{a}{5}} & \=& \frac{\i}{6}\left( A^R_{{a}}-e_{{a}}{}^{\mu}\partial_{\mu}\varphi \right)\,,\label{Ta554}\\
{\dot{V}_{a\,j}}{}^i & \=& {\cV_a}^i{}_{j}\,\,,\,\,\, {\dot{V}_{5\,j}}{}^i\= \phi^{-1}{\cY^{0\,i}{}_j} \,,\label{V54}\\
\dot{D}& \=& \frac{1}{4}\Big( \cD -\frac{1}{4}\phi^{-1}e^{a\mu} {D}_\mu ( e_a{}^{\nu }{D}_\nu \phi)-
\frac{3}{32}\phi^{-2}F(B)^{ab}F(B)_{ab} \nonumber\\
&& \qquad \qquad \qquad \qquad \qquad -\frac{3}{2}\dot T_{\dot{A}\dot{B}} 
\dot T^{\dot{A}\dot{B}}-\frac{1}{4}\phi^2 \dot{V}_{x^5 i }{}^j \dot{V}_{x^5 j}{}^i \Big)\,, \label{D54}
\eeqa
where
\be \label{phiBmap}
\phi \= 2\i e^{\varphi}X^0 \= -2\i e^{-\varphi} \overline{X}^0\,, \qquad\quad B_\mu = A^0_{\mu}\,.
\ee
The 5d supersymmetry parameters are:
\beqa 
\dot{\ve}^i & \= & {\rm e}^{\frac{1}{2}\varphi \gamma_5}\varepsilon_{4d}^i  \, \label{Qkslift},\\
\dot{\eta}^i & \= & \gamma_5\left(\i  {\rm e}^{-\frac{1}{2}\varphi \gamma_5}\eta_{4d}^i + 2 \dot{T}_{{a}{5}}\gamma^{a}  
\dot{\varepsilon}^i -\frac{\i}{8} \phi^{-1} (F(B)_{{a}{b}} -4\i \phi \dot{T}_{{a}{b}}\gamma_5 )   \gamma^{ab}\dot{\varepsilon}^i\right)\,. \label{Skslift}
\eeqa
The 5d vector multiplet is given in term of the 4d vector multiplet as:
\beqa
\dot\sigma^I &\=&  e^{ \varphi}X^I+  e^{- \varphi} \overline{X}^I \,, \label{gaugino54}\\
\dot{W}^I_{{a}} & \=& A^{I}_{{a}} \,,  \quad\quad \dot{W}^I_{{5}}\=-\i \left(e^{ \varphi}X^I-e^{- \varphi} \overline{X}^I\right) \label{W554} \,,\\
\dot{\Omega}^{I\,i}& \= & e^{\frac{1}{2}  \varphi \gamma_5} \lambda^{i~I} + \dot{W}^I_{{5}} \dot{\Psi}_{{5}}^{~i},\\
\dot{Y}^{I\,i}{}_j & \= &  -\half \cY^{I\,i}{}_j -\half \dot{W}^I_{{5}} {\dot{V}_{5\,j}}{}^i  \label{Y54}\,.
\eeqa
The 5d hyper scalar given in terms of the 4d hypermultiplet is
\beqa
\dot{A}_i{}^\alpha \= \phi^{1/2} \cA_i{}^\alpha \label{hyper54}\,.
\eeqa

\vskip 0.2cm

\ndt \textbf{Mapping~4d/5d classical backgrounds:} By the above 4d/5d map, the relation between the 4d 
AdS$_2 \times$S$^2$ backgrounds (\ref{backgroundE}) and 5d AdS$_3\times$ S$^2$ background 
in~\eqref{backgroundE5d} in KK coordinates becomes more manifest.
One important subtlety is about the choice of~$\varphi$ in~\eqref{phiBmap}. In the case of the Lorentzian 
4d/5d connection,~$\varphi$ is just a~$U(1)_R$ gauge parameter that fixes the gauge-redundant phase 
of~$X^0$ and~$\overline{X}^0$, making the $\phi$ automatically real. However, in the Euclidean case, 
the 4d theory has an~$SO(1,1)_R$ gauge symmetry instead of~$U(1)_R$, whereas the background values 
for $X^0$ and $\overline{X}^0$ have a relative phase coming from the choice of the parameter~$\omega$ 
and $\bar{\omega}$ and  value of the charge  $e^0$ and $p^0$. 
Therefore, unlike in the Lorentzian case, the value of~$\varphi$ is not a `gauge fixing' to kill the phase 
of~$X^0$ and~$\overline{X}^0$,  but rather a `choice' to cancel the phase of $X^0$ and $\overline{X}^0$. 
(By the~$SO(1,1)_R$ gauge redundancy and by the rule~\eqref{SO(1,1)}, we shift the~$\varphi$  to set 
the magnitude of $X^0$ and $\overline{X}^0$ to be same.)

Recalling the background value of $X^0$ and $\overline{X}^0$ as given in~\eqref{backgroundE}, where 
the~$\omega$ and~$\bar\omega$ are parametrized by $\alpha$ as in~\eqref{omChoices}, the value of 
the mapping parameter~$\varphi$ is determined to be 
\be \label{varphiconstraint}
\varphi^\pm(\alpha\,, e^0\,,p^0) = -\i \alpha \pm \i \frac{\pi}{2}  - \i \arctan\Bigl(\frac{p^0}{e^0} \Bigr)\,,
\ee
by the condition that~$\phi$ be real.
There remains an ambiguity of $\pm \pi /2$ that is related to an overall sign choice for~$\phi$. 
We now consider specific examples for two distinct choices of~$(e^0\,, p^0)$, keeping the choice 
of~$\alpha$ to be generic. These are:  
\be \label{varphi1}
\begin{split}
(1) \quad  (e^0,\,p^0) &\=  (e^0,\,0)  \,, \qquad \varphi^\pm_{1}(\alpha) \= -\i\alpha  \pm \i \pi /2\, \Rightarrow\,\phi \= \mp \frac{e^0}{\ell}\,, \\
(2) \quad (e^0,\,p^0) &\=  (0,\,p^0)\,, \qquad \varphi^\pm_{2}(\alpha) \= -\i(\alpha+\pi/2) \pm \i\pi /2  \Rightarrow\,\phi \= \mp \frac{p^0}{\ell}\,.
\end{split}
\ee
Here we see that, by the mapping parameter $\varphi^\pm$,  the background value of the lifted 5d field~$\phi$ 
is indeed real, but there is dependence on the choice~$\pm$.
We note that for both cases in~\eqref{varphi1} and, more generally, with any choice~\eqref{varphiconstraint}, 
all the lifted 5d fields are independent of the choice of phase $\omega\equiv\exp(\i\alpha)$ in the 4d 
background~\eqref{backgroundE}. 

The resulting 5d background fields are listed in Table~\ref{table:backgrounds}.
The~4d configuration with $(e^0,\,p^0,\,\varphi) =  (e^0,\,0,\,\varphi_1^\pm) $ as in (1) lifts to an  AdS$_3\times$S$^2$ 
background, while the one with $(e^0,\,p^0,\,\varphi) =  (0,\,p^0,\,\varphi_2^\pm) $ as in (2) lifts to an 
AdS$_2\times$S$^3$ background. For the latter case, the localization solutions were studied in~\cite{Gupta:2019xac}.
In both cases, the choice of the sign in~$\varphi^\pm$ gives the opposite sign for the background values 
of $\phi$, $\dot{T}_{\dot{A}\dot{B}}, \dot \sigma$ and hyper norm~$\dot \chi$. At the level of the 
Killing spinor equation (that we review in Appendix \ref{KSads3s2}), choosing either sign gives a set of 
Killing spinors corresponding, respectively, to the right- or left-moving supercharges in terms of the 
2d chiral $\cN=4$ super algebra.  
  \begin{table}
\renewcommand{\arraystretch}{1.5}
  \begin{center}
    \begin{tabular}{|c|c|}    
                \cline{2-2}
   \multicolumn{1}{c|}{}& $\varphi = \varphi_1^\pm$ \\
	\hline
	$(e^0,\,p^0) = (e^0, \,0)$ &  \makecell{ $ds^2 =$ AdS$_3 \times$S$^2$ \\ $\dot{T}_{34} 
	= \mp 1/(4 \ell)$ \\  $\dot{\sigma}^I = \mp p^I / \ell\,,$\\ $ \dot{F}^I_{3 4} 
	= p^I /\ell^2$\\$\dot{A}_{1,2}{}^{1,2}=\sqrt{\pm\frac{p^3}{3\ell^3}}$\\$S_{\text{bulk}} = \frac{p^3}{12 e^0}$ } \\
	\hline
\end{tabular}
\quad
\begin{tabular}{|c|c|}    

    		                \cline{1-1}
		                  $\varphi = \varphi_2^\pm$ &   \multicolumn{1}{c}{} \\
		                   \hline
 \makecell{$ds^2 =$ AdS$_2 \times$S$^3$ \\ $\dot{T}_{12} = \mp \i/(4 \ell)$ \\  $\dot{\sigma}^I = 
 \pm e^I / \ell\,, $\\$ \dot{F}^I_{1 2} = -\i e^I /\ell^2$\\ $\dot{A}_{1,2}{}^{1,2}=\sqrt{\mp\frac{e^3}{3\ell^3}}$\\ 
 $S_{\text{bulk}} =  \frac{e^3}{12 p^0}$ } & $(e^0,\,p^0) = (0, \,p^0)$   \\
\hline
          \end{tabular}
                \caption{The non-trivial 5d fields obtained by lifting the 4d backgrounds~(\ref{backgroundE}) 
                with~$(\omega,\,\overline{\omega})$ as given in~\eqref{omChoices} and with different choices 
                for~$(e^0,\,p^0)$ and~$\varphi$. For the choice of~$(e^0,\,p^0)$ on the left and right panel, the 
                4d hyper scalar that is lifted is determined by the~$\cD$-field equation constraint~\eqref{4dhyperconstraint} 
                as~${a_1{}^1 = a_2{}^2 = {1}/{\ell}\sqrt{-{p^3}/{3e^0}}}$ and~${a_1{}^1 = a_2{}^2 = {1}/{\ell}\sqrt{{e^3}/{3p^0}}}$ respectively. 
We also include the value for the finite piece of the bulk action~\eqref{bulk_action}.
      The field configurations on the right entry are solutions corresponding to the near-horizon of the supersymmetric Euclidean 5d black hole. 
      The field configurations on the left entry are the Euclidean AdS$_3 \times $S$^2$ solutions.}
       \protect{\label{table:backgrounds}}
  \end{center}
  \end{table}

Now, for our problem, the full specification of parameters to lift the Euclidean AdS$_2 \times$S$^2$ 
backgrounds~(\ref{backgroundE}) to the 5d KK frame~\eqref{backgroundE5d} is 
\be\label{e0p0varphi}
(e^0,\,p^0,\,\varphi) \=(-1,\,0,\, \varphi^+_1)\,,
\ee 
with identification~$e^I=\mu^I$ and~$\varphi^+_1$ given in~\eqref{varphi1}.
To relate Euclidean AdS$_2 \times$S$^2$ to the twisted torus~\eqref{TwistedFrame}, this lift is then followed 
by the following steps: taking the lifted 5d KK frame background~\eqref{backgroundE5d} with $Q$- Killing 
spinors~\eqref{4SMKSinKK}, one applies the local 
coordinate transformations~\eqref{KKcoords},~\eqref{twistedcoord}, the spinor Lorentz rotation in~\eqref{VKKtoGlobal} with~\eqref{GlobaltoKKSpin}, and finally one imposes the periodicity conditions~\eqref{Newperiodicity} with~$\Omega$ 
given in~\eqref{OmegaValue}.
In this procedure, only four of the eight $Q$- Killing spinors mapped from~\eqref{4SMKSinKK} are well-defined on the 
twisted torus, as expected.

\vskip 0.2cm

\ndt \textbf{Mapping 4d localization solution to the 5d twisted torus frame:}  Having identified the 
relevant 4d background, together with the correct mapping parameter~\eqref{e0p0varphi} that relates it to the 
5d twisted torus background~\eqref{TwistedFrame}, we now want to map the \textit{off-shell} localization 
solution of 4d supergravity on that background to the 5d localization solution around the twisted torus background.
The strategy for this mapping follows the same steps as the mapping of the backgrounds presented above. 
Here, we assume that phase factors in the quantum fluctuation of the scalars~$X^0$ and~$\overline{X}^0$ 
are appropriately cancelled by a fluctuating value of $\varphi$ around its value in \eqref{e0p0varphi}, such that 
it makes the quantum fluctuation of the 5d field~$\phi$~real.\footnote{Since we choose the reality condition 
for the fluctuation of $X^0$ and $\overline{X}^0$ to be complex conjugate to each other, as explained 
after~\eqref{4dreality}, and since this condition is the same as the condition in the Lorentzian theory, 
it appears there may be some $U(1)_R$ gauge symmetry hidden in the fluctuating field, and it may justify our assumption. 
}
It will turn out that for our off-shell localization solution, we can use the same value of $\varphi$ as was chosen in~\eqref{e0p0varphi}. 

Here, we summarize the steps as follows:
\begin{enumerate}
\item Start with the 4d localization manifold whose background is the Euclidean AdS$_2 \times $S$^2$ 
solution (\ref{backgroundE}) with $(e^0,\,p^0)=(-1,0)$. Since the result  does not depend on the choice 
of~$\alpha$ in~\eqref{omChoices}, without loss of generality we take $\alpha = \pi/2$ for convenience.
\item Apply the 4d/5d lift with the mapping parameter~$\varphi = \varphi_1^+(\pi/2)=0$ to obtain 5d localization 
solutions in the KK frame~\eqref{backgroundE5d} of Euclidean AdS$_3 \times $S$^2$. 
\item Transform these localization solutions to the twisted torus frame by applying the local coordinate 
maps~\eqref{KKcoords},~\eqref{twistedcoord}, the spinor Lorentz rotation in~\eqref{VKKtoGlobal} 
with~\eqref{GlobaltoKKSpin}, and finally imposing the periodicity conditions~\eqref{twistwithtau1}
with~$\Omega$ given in~\eqref{OmegaValue}. 
\end{enumerate}
Note that a consistent lift to the twisted torus requires that the lifted solutions respect 
the periodicities~\eqref{twistwithtau1}. As an example of an inconsistent lift, 
consider a scalar field fluctuation on AdS$_2 \times$S$^2$ with non-zero momentum on~$\chi$, 
which therefore has $2\pi$-periodicity in~$\chi$. 
Recalling that~$\chi=\psi - \i t_E$, we see that such a mode, lifted to 5d, does not respect the second periodicity 
condition in~\eqref{twistwithtau1}. As we discuss in the next section, the fields in the four-dimensional 
localization manifold depend only the radial coordinate~$\eta = 2\rho$ and therefore lift consistently  
to the 5d twisted torus.

\section{The lift of localization solutions on~AdS$_2 \! \times \!\text{S}^2$ to~$\Hthrees$ } \label{section:locresults}

In this section, we apply the lifting procedure to obtain localization solutions around the supersymmetric~$\Hthrees$   
background.
We find a set of solutions to the BPS equations parametrized by~$N_\text{v} +1$ 
real coordinates~$C^\cI$, $\cI=0, \dots, N_\text{v}$.
These coordinates are inherited from the 4d localization manifold, where each~$C^{\cI}$ parametrizes 
the off-shell solution for the~$\cI^{th}$ vector multiplet.
In the 4d AdS$_2 \times$S$^2$ problem, the boundary conditions fix all the fields to their attractor values at infinity.
The localization solution consists of the scalar fields~$X^\cI$ going off-shell in the interior, with 
a radially-decaying shape that is fixed by supersymmetry. The parameter~$C^\cI$ labels the size of deviation at the origin.
In 5d, the~$C^{I}$,~$I=1,\,\cdots,\,N_\text{v}$ parametrize the size of the off-shell solution in the 
vector multiplet, and~$C^0$ parametrizes a certain excitation of the Weyl multiplet.
Here, we have an~AdS$_3\times$S$^2$ background, where one leg of the gauge 
field~($W_{{z'}}$) is fixed at infinity to its on-shell value while the other ($W_{\overline{z'}}$) 
is free to fluctuate, as we described in Section~\ref{sec:Semiclassics}.
The parameter~$C^I$ labels the deviation of both~$W^I_{z'}$ and~$W^I_{\overline{z}'}$ 
from their on-shell value at the origin as well as the boundary fluctuation of~$W^I_{\overline{z}'}$.
The precise solutions are presented in~(\ref{OffShellE5dGlobal}--\ref{DGlobal}) for the Weyl multiplet, and 
in~(\ref{sigmaGlobal}--\ref{D3Global}) 
for the vector multiplets. The hypermultiplet also fluctuates, and the solution is given in~\eqref{hyperGlobal}.

\vskip 0.2cm
\ndt
\textbf{4d localization solutions:}

\ndt
The most general solution in 4d around the~AdS$_2\times$S$^2$ background is parametrized by one real 
parameter in each vector multiplet and one real parameter in the Weyl multiplet, before fixing the gauge for local 
scale transformations~\cite{Gupta:2012cy}.
The gauge can be chosen so that there is no off-shell fluctuations in the Weyl multiplet \cite{Dabholkar:2010uh}.
The off-shell solution in the vector multiplets takes the following form:
\beqa \label{4dlocsolution}
 X^{\mathcal{I}}  && \=  \frac{\i}{2\ell}\,\left(e^{\mathcal{I}}+\i\, p^{\mathcal{I}}  +\frac{C^{\mathcal{I}}}{\cosh\eta}\right) \quad, \quad
 \overline{X}^{\mathcal{I}}\= -\frac{{\i}}{2\ell}\,\left(e^{\mathcal{I}}-\i \, p^{\mathcal{I}}  +\frac{C^\cI}{\cosh\eta}\right) \,, \label{4dvecsolX}\\
 A^{{\mathcal{I}}} && \= - \i \,e^{\mathcal{I}}(\cosh \eta -1)\,d\chi  -p^{\mathcal{I}} \cos\theta\,d\phi \,, \label{4dvecsolW}\\
{\cY^{\cI \,1}}_1 && \= \cY^\cI_{12} \= \frac{-C^{\mathcal{I}}}{\R^2 \cosh^2\eta}\,, \label{4dvecsolY}
\eeqa
where we use~$(\omega(\pi/2),\,\overline{\omega}(\pi/2)) = ( 4\i /\ell,\, - 4\i / \ell)$.
The $C^{\mathcal{I}}$ parametrize the off-shell fluctuations around the background \eqref{backgroundE}. 
\vskip 0.2cm
\ndt
\textbf{Lift to the Weyl multiplet:} 

\ndt
For the lift to the Weyl multiplet, the relevant fields of the 4d localization solution \eqref{4dlocsolution} are 
those of the KK vector multiplet~${\mathcal{I}}=0$.
Using~\eqref{phiBmap}, we first obtain the off-shell values for the KK scalar and one-form:
\be
\phi  \=   \frac{1}{\ell}\left(1 - \frac{C^0}{\cosh \eta}\right) \,,\qquad B_\chi  \=  \i (\cosh \eta -1)\,.
\ee
It is useful to define the function
\be
\phi(x) \, \defeq \, 1-\frac{C^0}{\cosh x}\,.
\ee
Now, using the lifting equations~(\ref{E54} - \ref{D54}) with $(e^0,\,p^0) = (-1,\,0)$ and~$\varphi  = 0$, 
we obtain the full Weyl multiplet configuration in the KK frame.
After applying the coordinate maps~\eqref{KKcoords} and~\eqref{twistedcoord} to the twisted torus frame, the non-trival fields are:
\beqa 
{E_{M'}}^A & \= & \ell
\begin{pmatrix}
2 & 0 && 0 && 0 & 0\\
0 &\sinh \rho \left(1 + \frac{1- C^0}{\gmr} \right)  && 0 && 0 &\i  \cosh \rho \left( 1- \frac{1-C^0}{\gmr}\right)\\
0 & 0 && 1 && 0 & 0 \\
0 & 0 &\,\,& 0 &\,\,&\sin \theta & 0\\
0 &-\i\sinh \rho  \left(  1- \frac{1+C^0}{\gmr}\right) && 0 && \i  
\Omega \sin \theta &\,\,\,\cosh \rho \left(1 + \frac{1+C^0}{\gmr}\right)
\end{pmatrix}\,, \label{OffShellE5dGlobal}\\
T_{\theta \phi'}& \= &\ell \sin \theta\frac{1}{12}\left(\frac{1}{\gmr}-4\right)\,, \qquad T_{\theta {t_E}'}
\=\i \ell \sin \theta\frac{\Omega }{12}\left(\frac{1}{\gmr}-4\right)\,, \\
{V}_\psi & \= &\i \left( 1-\frac{1}{\cosh2\rho}\right)\frac{\gmr -1}{\phi^2(2\rho)}\btau_3\,,~~ {V}_{t_E'} \=\left( 1+\frac{1}{\cosh2\rho}\right)\frac{\gmr -1}{\phi^2(2\rho)}\btau_3\,,
\\
D 
&\=& \frac{ 1}{8 \ell^2\gmr^2}\left( (1- \gmr )\frac{\sinh^2 2\rho}{\cosh^2 2\rho}- \frac{2}{3}(1-\gmr)^2  \right)\,.\label{DGlobal}
\eeqa
The line-element corresponding to the vielbein~\eqref{OffShellE5dGlobal} is:
\be
\begin{split}
ds^2 
\=& 4\R^2 d\rho^2 +4 \ell^2 \left( \sinh^2 \rho  -\sinh^4\rho  \left(    \frac{1}{\phi^2(2\rho)}-1\right) \right)d\psi^2
\\
 &+\R^2 \Bigl(d\theta^2 + \sin^2\theta  \bigl(d\phi' + \i \Omega dt'_E\bigr)^2 \Bigr) + 2 \i \ell^2 \sinh^2 2\rho \left( \frac{1}{\phi^2(2\rho)}-1 \right) d\psi d{t'_E} 
 \\
& +4\ell^2\left(  \cosh^2\rho + \cosh^4\rho \left(\frac{1}{\phi^2(2\rho)}-1  \right)  \right)d{t_E'}^2
\,.
\end{split}\ee
Here, recall from Section~\ref{sec:twistedbackgnd} that~$\Omega = 1+ \i \tau_1/\tau_2$ in the twisted torus frame.

It remains to apply the lift to the $Q$- and $S$-Killing spinors.
In principle, off-shell fluctuations in the bosonic fields of the Weyl multiplet may induce off-shell fluctuations in 
the 5d Killing spinors such that the BPS equations of the multiplet remain solved.
Note however that the 4d Weyl multiplet in the 4d localization solution does not fluctuate, and so the 4d $Q$- 
and $S$- Killing spinors that we lift are just those of the 4d background, namely the eight spinors $\ve^i_{4d}(\pi/2)$, 
given explicitly in~\eqref{8SMKSinKK}, and~$\eta^i_{4d}(\pi/2)=0$ (recall we have fixed~$\alpha = \pi/2$).
Further note that the lifting equation~\eqref{Qkslift} for the 5d $Q$- spinors only involves the 4d $Q$- spinors (which are on-shell).
We conclude that the lift of the $Q$- spinors is unchanged from the on-shell case, i.e.~we obtain, in the twisted torus frame, 
the four well-defined on-shell $Q$ -spinors~$\ve_{(a)}$,~$a=1,2,3,4$, as given in~\eqref{4TwistedSMKS}. 
In contrast, the lifting equation~\eqref{Skslift} of the $S$- spinors~$\eta_{4d}$ involves bosonic 5d fields which do fluctuate.
The 5d $S$- spinors, which are zero on-shell, therefore acquire a non-zero value off-shell.
In the twisted frame, we obtain four well-defined $S$- spinors~$\eta_{(a)}$, associated with the four~$Q$- spinors~$\ve_{(a)}$.
The one associated to the localization supercharge~$\ve_{(1)}$ has value
\beqa
\eta^1_{(1)} & \= & -\frac{\frac{C^0}{ \cosh \left( 2\rho \right) }\,{\rm e}^{\frac{\i}{2} (\psi +\phi' +\i(\Omega-1)t_E')}}{3 \sqrt{2}\ell \gmr}
\begin{pmatrix}
{\cos{\frac{\theta}{2}}\,\cosh{\frac{\rho}{2}}}\\
-{\sin{\frac{\theta}{2}}\,\cosh{\frac{\rho}{2}}}\\
-{\cos{\frac{\theta}{2}}\,\sinh{\frac{\rho}{2}}}\\
{\sin{\frac{\theta}{2}}\,\sinh{\frac{\rho}{2}}}
\end{pmatrix}\,,\\
 \eta^2_{(1)}  & \= & -\frac{ \i \frac{C^0}{\cosh  \left( 2\rho \right)}\,{\rm e}^{-\frac{\i}{2} (\psi +\phi' +\i(\Omega-1)t_E')}}{3 \sqrt{2}\ell\,\gmr}
 \begin{pmatrix}
{\sin{\frac{\theta}{2}}\,\sinh{\frac{\rho}{2}}}\\
{\cos{\frac{\theta}{2}}\,\sinh{\frac{\rho}{2}}}\\
-{\sin{\frac{\theta}{2}}\,\cosh{\frac{\rho}{2}}}\\
-{\cos{\frac{\theta}{2}}\,\cosh{\frac{\rho}{2}}}
\end{pmatrix}\,.
\eeqa

\bigskip

\ndt \textbf{Lift of the vector multiplet:} 

\ndt
The relevant 4d fields are those of \eqref{4dlocsolution} with~${\mathcal{I}}=I$.
Using the lifting equations~(\ref{gaugino54} - \ref{Y54}) followed by the coordinate 
transformations~\eqref{KKcoords} and~\eqref{twistedcoord}, we obtain the following 
non-trivial fields of the vector multiplet configuration in the twisted torus frame:
\beqa
\sigma^I  &&\= -\frac{p^I}{\ell}\,, \qquad \qquad W_{\phi'}^I  \= -p^I \cos \theta\,, \label{sigmaGlobal}\\
W_\psi^I && \= \frac{ 2\i \left({C^I/\mu^I} + C^0\right)\frac{\sinh^2(\rho)}{\cosh 2 \rho}}{\gmr}\mu^I\,,
\eeqa
\beqa
W^I_{t_E'} && \=-\i p^I \Omega \cos \theta + \frac{\frac{C^I/\mu^I-C^0}{\cosh 2\rho}+C^I/\mu^I+C^0+2 }{\gmr}\mu^I\,,\\
Y^I_{12} && \= \frac{1}{2\ell^2\gmr}\frac{C^I/\mu^I+C^0 }{\cosh^2 2\rho} \mu^I \label{D3Global}\,.
\eeqa

\bigskip

\ndt \textbf{Lift of the hypermultiplet:} 

\ndt
Finally, the lift for the hypermultiplet~\eqref{hyper54} gives the following non-trivial components for the off-shell hyper scalar:
\be \label{hyperGlobal}
{A}_1{}^1 \= {A}_2{}^2 \= \left(\frac{\gmr}{\ell}\right)^{1/2} \sqrt{\frac{p^3}{3\ell^3}}\,.
\ee

\medskip

To summarize, the field configuration of the Weyl multiplet~\eqref{OffShellE5dGlobal}--\eqref{DGlobal}, 
the vector multiplet \eqref{sigmaGlobal}--\eqref{D3Global}, and the hypermultiplet~\eqref{hyperGlobal} 
are the 5d localization solutions.
These configurations are off-shell fixed-points of the variations generated by the supercharge~$\cQ$ 
given in~\eqref{defCQ}, around the supersymmetric $\Hthrees$ given in~\eqref{TwistedFrame}.

\section*{Acknowledgements}
We would like to thank Satoshi Nawata and Valentin Reys for useful discussions and comments. 
This work is supported by the ERC Consolidator Grant N.~681908, ``Quantum black holes: A microscopic 
window into the microstructure of gravity'', and by the STFC grant ST/P000258/1. 
A.C.~is supported by the STFC studentship ST/S505468/1.
IJ is supported by an appointment to the JRG Program at the APCTP through the Science and Technology 
Promotion Fund and Lottery Fund of the Korean Government, by the Korean Local 
Governments - Gyeongsangbuk-do Province and Pohang City, and  by the National 
Research Foundation of Korea(NRF) grant funded by the Korea government(MSIT) (No. 2021R1F1A1048531).


\newpage
\appendix

\section{Notations and conventions  \label{sec:spinors}}

We summarize the various index notations in Table~\ref{table:notations}.
\begin{table}[h]
\begin{center}
\renewcommand{\arraystretch}{1.5}
    \begin{tabular}{|c|c|c|c|}   
    \hline
    \textbf{Index} & \textbf{Range} & \textbf{Description}\\
      \hline
$M,\,N,\,\cdots$ &  $(\rho,\,\psi,\,\theta,\,\phi,\,t_E)$ & 5d cylinder coordinates\\
      \hline
$M',\,N',\,\cdots$ & $(\rho,\,\psi,\,\theta,\,\phi',\,t_E')$ &5d twisted torus coordinates \\
      \hline
      $A,\,B,\,\cdots$ & $(\hat\rho,\,\hat\psi,\,\hat\theta,\,\hat\phi,\,\hat{t_E})$  & tangent frame for cylinder and twisted torus coordinates\\
   \hline
$\dot{M},\,\dot{N},\,\cdots$ & $(\eta,\,\chi,\,\theta,\,\phi,\,x^5)$& 5d Kaluza-Klein coordinates \\
      \hline
$\dot{A},\,\dot{B},\,\cdots$ & $(1,\,2,\,3,\,4,\,5)$ & tangent frame for Kaluza-Klein coordinates\\
      \hline
$\mu,\,\nu,\,\cdots$ & $(\eta,\,\chi,\,\theta,\,\phi)$ &  Euclidean AdS$_2\times$S$^2$ coordinates \\
\hline
$a,\,b,\,\cdots$ &$(1,2,3,4)$& tangent frame for Euclidean AdS$_2\times$S$^2$ coordinates\\
      \hline
$\alpha,\,\beta,\,\cdots$ &   $(z,\overline{z})$ & thermal AdS$_3$ boundary coordinates\\
\hline
$i,\,j,\,\cdots$ & $(1,\,2)$ or $(+,\,-)$ &  Fundamental SU(2) \\
\hline 
      \end{tabular}
      \caption{Summary of index notation.}
      \label{table:notations}
   \end{center}
  \end{table}

\ndt
\textbf{Spinors and gamma matrices:}

We denote a basis for the the $d$-dimensional Clifford algebra as
\be \label{cliffordbasis}
\left\{ \Gamma \= \mathbb{I},\,\gamma^{A_1},\,\gamma^{A_1 A_2},\cdots \gamma^{A_1 A_2 \cdots A_d}\right\}\,,
\ee
where:
\be
\gamma^{A_1 \cdots A_k} \= \gamma^{[A_1}\cdots  \gamma^{A_k]}\,.
\ee
In five dimension with Lorentzian signature, a consistent choice of gamma matrix satisfies the following relations:
\begin{equation}\label{5dgammaLor}
\begin{array}{llll}
\gamma_{A}^{\dagger}\=-A\gamma_{A}A^{-1}\,,~&~A\=\gamma_{0}\,,~&~A^{\dagger}\=A^{-1}\=-\gamma_{0}\,,~&\\
\gamma_{A}^{T}\={\cal{C}}\gamma_{A}{\cal{C}}^{-1}\,,~&~{\cal{C}}^{T}\=-{\cal{C}}\,,~&~{\cal{C}}^{\dagger}={\cal{C}}^{-1}\,,~&\\
\gamma_{A}^{*}\=-B\gamma_{A}B^{-1}\,,~&~B^{T}\={\cal{C}}A^{-1}\,,~&~B^{\dagger}\=B^{-1}\,,~&~B^{*}B\=-1\,.
\end{array}\end{equation}
This is followed by the property, regarding the charge conjugation matrix $\cC$, 
\begin{equation}
({\cal{C}}\Gamma^{(r)})^{T}= - (-)^{r(r-1)/2}{\cal{C}}\Gamma^{(r)}\,,
\end{equation}
where~$\Gamma^{(r)}$ is a matrix of the set~\eqref{cliffordbasis} with rank~$r$.
Due to the property of the charge conjugation matrix, we can use the spinor representation satisfying the symplectic-Majorana condition
\be\label{SMLor}
\bar{\psi}_ i \= (\psi^i)^\dagger \gamma_0\,,
\ee
where~$i$ is an SU(2)$_R$ index, and where $\bar{\psi}_i$ is the symplectic-Majorana conjugate of $\psi^i$, defined as
\be\label{DefSM} 
\bar{\psi}_i \defeq \varepsilon_{ij}(\psi^j)^T \mathcal{C}\,,
\ee
with $\varepsilon_{ij}$ being the SU(2) symplectic metric $\varepsilon_{12} = -\varepsilon_{21} = 1$.\\

The five-dimensional Euclidean case is obtained by the Wick rotation of time direction~$x^0$, using the redefinition:
$
x^0 = -\i x^5\,.
$
This consistently redefines the $0^{th}$ gamma matrix as the~$5$-th directional one as
$
\gamma_0 = \i \gamma_5\,.
$
The relations on the Lorentizan gamma matrices~\eqref{5dgammaLor} then become, for the Euclidean case:
\begin{equation}\label{5dgammaEucl}
\begin{array}{lll}
\gamma_{A}^{\dagger}\=\gamma_{A}~&~\\
\gamma_{A}^{*}\=\gamma_{A}^{T}\= {\cal{C}}\gamma_{A}{\cal{C}}^{-1}\,,~&~{\cal{C}}^{\dagger}=
{\cal{C}}^{-1}\,,~&~{\cal{C}}^{T}\=-{\cal{C}}\,~\Leftrightarrow~{\cal{C}}^{*}{\cal{C}}\=-1\,,\\
\end{array}\end{equation} 
with charge-conjugation matrix property:
\begin{equation} \label{Ctranspose}
({\cal{C}}\Gamma^{(r)})^{T}\= - (-)^{r(r-1)/2}{\cal{C}}\Gamma^{(r)}\,.
\end{equation} 
In the main text, we often consider Lorentz scalars of the type
\be
\bar{\lambda}_i\,\Gamma^{(r)}\epsilon^j \,.
\ee
For two Grassman even spinors~$\epsilon^i$, $\lambda^j$, the property~\eqref{Ctranspose} leads to the 
following Majorana-flip relations:
\be \label{flip}
\bar{\lambda}_i\,\Gamma^{(r)}\epsilon^j \=  (-)^{r(r-1)/2}\left( \delta^j_i \,\bar{\epsilon}_k\,
\Gamma^{(r)}\lambda^k-\bar{\epsilon}_i\,\Gamma^{(r)}\lambda^j\right)\,.
\ee
Note some useful consequences of~\eqref{flip} for~$\lambda = \epsilon$:
\beqa
\bar{\epsilon}_i\, \epsilon^j &&\=  \half  \left(\bar{\epsilon}_k\,\epsilon^k\right)\,\delta^j_i\,,\\
\bar{\epsilon}_i\,\gamma^{A}\epsilon^j &&\=  \half  \left(\bar{\epsilon}_k\,\gamma^{A}\epsilon^k\right)\,\delta^j_i\,, \label{KMid}\\
\bar{\epsilon}_k\,\gamma^{AB}\epsilon^k  &&\= 0\,, \qquad 
\bar{\epsilon}_k\,\gamma^{ABC}\epsilon^k  \=  0\,,
\eeqa
where we used~$r = 0,1,2,3$ respectively.
The spinors in the Euclidean theory can also be chosen to be symplectic-Majorana, but differently from \eqref{SMLor}, satisfying
\be\label{SMEucl}
\bar{\psi}_ i \= (\psi^i)^\dagger\,,
\ee
with the same definition of the symmplectic Majorana conjugate $\overline{\psi}_i$ as \eqref{DefSM}. 
However, we note that, as is commented in the begining of section~\ref{5dSUGRA}, we does not 
impose \eqref{SMEucl} for quantum theory.

\section{Supersymmetry transformations in Euclidean 5d supergravity \label{app:susytrans}}
\begin{table}[H]
\begin{center}
\renewcommand{\arraystretch}{1.5}
    \begin{tabular}{|c||c|c|c|}    
      \hline
      Weyl  & ${E_M}^A \, ,\,\Psi^i_M \, ,\, b_M \, ,\, {V_{M,\,i}}^j \, ,\,T_{MN}\, ,\,D \, ,\,\chi^i$ \\
      \hline
      Vector   & $\sigma^I \, ,\,W^I_M \, ,\, \Omega^{I\,i} \, ,\, Y^I_{ij} $  \\
      \hline
       Hyper & $A_i{}^\alpha \, ,\,\zeta^\alpha$ \\
      \hline
      \hline
      SUSY parameters & $\epsilon^i \, ,\,\eta^i$\\
      \hline
      \end{tabular}
      \caption{Independent fields of the supersymmetric multiplets and $Q$, $S$-supersymmetry parameters 
      in five-dimensional $\mathcal{N}=1$ conformal supergravity.}
   \label{table:fieldcontent}
   \end{center}
  \end{table}
Up to higher order in fermions, the infinitesimal $Q$- and $S$-supersymmetry transformations 
on the spinor fields of the theory are as follows:
\be  \label{KSequation1}
\begin{split}
\delta \Psi^i_M &\=  2\mathcal{D}_M \epsilon^i +\frac{i}{2}T_{AB}(3\gamma^{AB}\gamma_M-\gamma_M\gamma^{AB})
\epsilon^i-\i \gamma_M \eta^i  \,,\\
\delta \chi^i & \=  \half  \epsilon^i D +\frac{1}{64} R_{MN j}{}^i(V) \gamma^{MN}\epsilon^j +
\frac{3\i}{64} (3\gamma^{AB} \gamma^C+\gamma^C \gamma^{AB} )
\epsilon^i{D}_C T_{AB}  \\
& \qquad -\frac{3}{16}T_{AB}T_{CD}\gamma^{ABCD}\epsilon^i +\frac{3}{16}T_{AB}\gamma^{AB}\eta^i\,, \\
\delta \Omega^i & \=  -\frac{1}{2}(F_{AB}-4 \sigma T_{AB})\gamma^{AB}\epsilon^i-\i\gamma^A\epsilon^i{D}_A\sigma  -2 
\varepsilon_{jk}Y^{ij}\epsilon^k + \sigma \eta^i \,,  \\
\delta \zeta^\alpha & \= -\i \gamma^A\epsilon^i{D}_A {A_i}^\alpha +\frac{3}{2}{A_i}^\alpha \eta^i  \,. 
\end{split}
\ee
where  the curvature~$R_{MN i}{}^j(V)$ is given by:
\be \label{Vcurv}
R_{MN i}{}^j(V) \= 2 \, \partial_{[M} {{{{V_{N]}}}}_i}^j-2{{V_{[M}}_i}^k{{V_{N]}}_k}^j\,.
\ee
  
The relevant supercovariant derivatives acting on each field are covariant with respect to all 
bosonic gauge symmetries except conformal boosts:
\be\label{covariant_derivatives}
\begin{split}
 {D}_M \, \epsilon^i  & \= \left(\partial_M-\frac{1}{4}\,{\omega_M}^{AB}\,
 \gamma_{AB}+\frac{1}{2}b_M\right)\epsilon^i+\frac{1}{2}{{V_M}_j}^i \, \epsilon^j \,,\\
D_M T_{AB} & \= (\partial_M - b_M) T_{AB}   - \omega_{MA}{}^{C}T_{CB} -\omega_{MB}{}^CT_{AC}\,,\\
 D_M \, \sigma^I &\= \left(\partial_M - b_M\right)\,\sigma^I \,,\\
D_M \, {A_i}^\alpha &\=  \left(\partial_M   -\frac{3}{2}b_M \right){A_{i}} ^\alpha-\frac{1}{2} {{V}_{M i}{}^j} {A_j}^\alpha\,.
\end{split}
\ee

\section{Killing spinors on AdS$_3$ and S$^2$ \label{sec: killing_spinors}}\label{KSads3s2}

In this appendix, we present solution of the Killing spinor equation~\eqref{KSE} on the AdS$_3\times$S$^2$  
background given in~\eqref{metric} and \eqref{TMN}.  Here, let us decompose 
the spacetime and local indices into 
those for $3+2$ dimensions as~$M =\{\mu \,, \text{m}\}$ and $A=\{a\, ,\ta \}$.  Then the Killing spinor 
equation \eqref{KSE} splits as
\be
\cD_\mu \epsilon^i \=  s \frac{\i}{4 \R }\gamma^{\hat\theta \hat\phi}\gamma_\mu \epsilon^i\,,\quad \quad
\cD_\tm \epsilon^i =  s \frac{\i}{2\R} \gamma^{\hat\theta\hat\phi}\gamma_\tm \epsilon^i\,,
\ee
where we inserted the sign factor $s=\pm 1$ to keep track of the choice of the background value of $T_{MN}$; $s= +1$ 
is  for our background value of $T_{MN}$ in~\eqref{TMN}, and $s=-1$ is  for another background value by 
changing $T_{MN}\rightarrow - T_{MN}$ from the~\eqref{TMN} (which involves changing $\sigma \rightarrow -\sigma$ 
from \eqref{vector_field_content} by the BPS equation of vector multiplet).  
Note that,  since the background metric \eqref{metric} is direct product of 3 and 2 dimensions, the spin connection 
is also well separated as~$-\frac{1}{4}\omega_\mu^{AB}\gamma_{AB}=-\frac{1}{4}\omega_\mu^{ab}\gamma_{ab} $ 
and~$-\frac{1}{4}\omega_\tm^{AB}\gamma_{AB}=-\frac{1}{4}\omega_\tm^{\ta\tb}\gamma_{\ta\tb} $.  
This can be seen explicitly by noting that the non-zero spin connection components are
\be\ba{lll}\label{spinconnection5dLor}
\omega_t^{\hat{t} \hat\rho}= -\sinh\rho\,, \quad \quad &\omega_\psi^{\hat\rho\hat\psi}= 
\cosh\rho\,,\quad\quad &\omega_\phi^{\hat\theta\hat\phi}=\cos\theta\,.
\ea\ee 

We now decompose the spinor as
\be
\epsilon^i \= \epsilon^i_{AdS_3}\otimes \epsilon^i_{S^2}\,,
\ee
and take the following decomposition  for the gamma matrices
\be\ba{llll}\label{gamma5dLor}
  \gamma_{\hat{t}}\= \bsigma_0 \otimes \btau_3\,,&
    \quad \gamma_{\hat\rho}\= 
    \bsigma_1 \otimes \btau_3\,,&
     \quad \gamma_{\hat\psi}\=
   \bsigma_2 \otimes \btau_3 \,,&
    \quad  \gamma_{\hat{\theta}}\=
        \mathbb{I} \otimes \btau_1\,,
   \quad \gamma_{\hat\phi}=\mathbb{I} \otimes \btau_2\,,
\ea\ee
where~$\btau_a$, $a=1,2,3$, denotes the Pauli sigma matrix and~$\bsigma_a$ 
with~$a=0,1,2$ denotes the~3 dimensional gamma matrix.  
Here we choose~$\bsigma_0 = -\bsigma_1\bsigma_2$ 
such that 
$
\gamma_{\hat{t}\hat{\rho}  \hat{\psi} \hat{\theta}\hat{\phi}} \=\i\,
$ 
for our convention. The charge conjugation matrix can also be set to
\be\label{chargeconjugation}
\cC \= -\i \bsigma_2 \otimes \btau_1\,,
\ee
such that the gamma matrix relation \eqref{5dgammaLor} is satisfied.
With this splitting of spinors and gamma matrices, we arrive at the Killing spinor equations 
for AdS$_3$ and S$^2$ with radii~$2 \ell$ and~$\ell$ respectively :
\be \label{KS_split}
 0\=\left(\cD_\mu \epsilon^i_{AdS_3} +s \frac{1}{4\R} \bsigma_\mu \epsilon^i_{AdS_3}\right)\otimes 
 \epsilon^i_{S^2}\quad , \quad 
 0=\epsilon^i_{AdS_3}\otimes \left(\cD_\tm \epsilon^i_{S^2} + s\frac{1}{2\R} \btau_3\btau_\tm \epsilon^i_{S^2}\right)\,.
\ee
The general solutions of these equations are well known~\cite{Lu:1998nu}, and the solutions are given by 
\beqa
\epsilon_{AdS_3}&\=& {\rm e}^{-s\frac{1}{2}\bsigma_1 \rho}{\rm e}^{- s \frac{1}{2} 
\bsigma_0 t}{\rme}^{\frac{1}{2} \bsigma_{12} \psi}A\,, \\
\epsilon_{S^2}&\=& {\rm e}^{-s \i  \frac{1}{2}\btau_2 \theta}{\rm e}^{ \i \frac{1}{2}\btau_3 \phi}B\,,
\eeqa
where~$A$ and~$B$ are constant two-component complex  spinors. 

Let us write down the Killing spinor explicitly. We set the sign factor $s=1$, denote the chiral and anti-chiral 
component of the constant spinors as~$A_\pm $ and~$B_\pm$, and choose the 3 dimensional gamma matrix representation as 
\be\label{sigma3d}
\bsigma_a \=(-\i \, \btau_3 \,,\btau_1\,,\btau_2)\,.
\ee
Then we can rewrite the solutions as
\beqa
\epsilon_{\text{AdS}_3}&\=& A_+ \epsilon^{+}_{\text{AdS}} + A_- \epsilon^-_{\text{AdS}}\,,~~~~\quad
\epsilon_{\text{S}^2}\= B_+ \epsilon^{+}_{\text{S}^2} + B_- \epsilon^-_{\text{S}^2}\,,
\eeqa
where
\beqa\label{KSads3pmv2}
&&\epsilon^{+}_{\text{AdS}_3}\= {\rm e}^{\frac{\i}{2}(t +\psi)}\Biggl( \ba{c}\cosh \frac{\rho}{2} \\ -\sinh\frac{\rho}{2}\ea\Biggr)\,, 
~~~~~ \epsilon^{-}_{\text{AdS}_3}\= {\rm e}^{-\frac{\i}{2}(t +\psi)}
\Biggl( \ba{c}-\sinh \frac{\rho}{2} \\ \cosh\frac{\rho}{2}\ea\Biggr)\,, \\
\label{KSs2pm}
&&\epsilon^{+}_{\text{S}^2}\= {\rm e}^{\frac{\i}{2}\phi}\Biggl( \ba{c}\cos \frac{\theta}{2} \\ \sin\frac{\theta}{2}\ea\Biggr)\,, 
~~~~~~~~~~~~~~~\epsilon^{-}_{\text{S}^2}\= {\rm e}^{-\frac{\i}{2}\phi}\Biggl( \ba{c}-\sin \frac{\theta}{2} \\ \cos\frac{\theta}{2}\ea\Biggr)\,.
\eeqa
By direct product of  the  Killing spinors \eqref{KSads3pmv2} and those of~\eqref{KSs2pm}, we obtain 
four complex basis of Killing spinors as \eqref{4complexKS}, or 8 pairs of symplectic Majorana spinors as in~\eqref{8SMKS}. 

Note that the effect of  the different  sign $s$  is to flip  the  sign of both $\rho$ and $t$ in the Killing spinors. 
We also note that  in odd dimensions there are two inequivalent representations of gamma matrix. 
For instance, we can also choose $\bsigma_a = (+\i \btau_3, \btau_1, \btau_2)$ instead of~\eqref{sigma3d}. 
Then this is equivalent to the changing  the sign of $t$ in the Killing spinors.

\section{Global symmetry generators of AdS$_3$ and S$^2$}\label{isogenerator}
The global AdS$_3$  geometry, in coordinates given in~\eqref{metric}, has isometries generated by the following Killing vectors,
\be\ba{lll}\label{SL2Rrep}
\bar\ell_- &=&\frac{1}{2}\left[ \tanh\rho\,  {\rm e}^{-\i (t - \psi)} \partial_t - 
\coth\rho \,{\rm e}^{-\i (t-\psi)} \partial_\psi +\i {\rm e}^{-\i (t-\psi)}\partial_\rho \right]\,,\\
\bar\ell_0 &=& -\frac{\i}{2} (\partial_t - \partial_\psi)\,,\\
\bar\ell_+ &=& -\frac{1}{2}\left[ \tanh\rho\,  {\rm e}^{\i (t - \psi)} \partial_t - 
\coth\rho \,{\rm e}^{\i (t-\psi)} \partial_\psi -\i {\rm e}^{\i (t-\psi)}\partial_\rho \right]\,,\\
\ell_- &=&\frac{1}{2}\left[ \tanh\rho\,  {\rm e}^{-\i (t + \psi)} \partial_t + 
\coth\rho \,{\rm e}^{-\i (t+\psi)} \partial_\psi +\i {\rm e}^{-\i (t+\psi)}\partial_\rho \right]\,,\\
\ell_0 &=& -\frac{\i}{2} (\partial_t + \partial_\psi)\,,\\
\ell_+ &=& -\frac{1}{2}\left[ \tanh\rho\,  {\rm e}^{\i (t + \psi)} \partial_t + 
\coth\rho \,{\rm e}^{\i (t+\psi)} \partial_\psi -\i {\rm e}^{\i (t+\psi)}\partial_\rho \right]\,.
\ea\ee
They form the $SL(2,\mathbb{R})_L\times SL(2,\mathbb{R})_R$ algebra through the Lie bracket:
\be\ba{ll}
\left[\bar{\ell}_0 \,, \bar{\ell}_\pm \right]_{\text{Lie}} =\pm \bar{\ell}_\pm\,,~~~~& \left[ \bar{\ell}_+\,, 
\bar{\ell}_- \right]_{\text{Lie}}= -2 \bar{\ell}_0\,, \\
\left[ \ell_0 \,, \ell_\pm \right]_{\text{Lie}}  =\pm \ell_\pm\,,~~~~& \left[ \ell_+\,, \ell_- \right]_{\text{Lie}}= -2 \ell_0\,.
\ea\ee
The S$^2$ geometry, in coordinates given in~\eqref{metric}, has the isometries  generated by the following Killing vectors,
\be\ba{lll}\label{SO3rep}
{j}_1 &=& \i (\sin\phi \partial_\theta + \cos\phi \cot\theta \partial_\phi)\,,\\
j_2 &=& -\i (\cos\phi \partial_\theta - \sin\phi \cot\theta \partial_\phi)\,,\\
j_3 &=& -\i \partial_\phi\,,
\ea\ee
which satisfy the $SO(3)$ algebra
\be
[j_i \,, j_j]_{\text{Lie}} = \i \epsilon_{ijk}j_k\,.
\ee
\bigskip

The bosonic sector of the supersymmetry algebra of AdS$_3 \times$S$^2$ presented in Section~\ref{LorentzianKSandAlgebra}, contains $SL(2,\mathbb{R})_R$ and the $SO(3)$ symmetry generators, acting on all the fields of 5d supergravity.
Their representations  as variations on fields are given by the combination of the differential operators
presented in~\eqref{SL2Rrep},~\eqref{SO3rep}
with the corresponding local Lorentz transformation given as follows:\footnote{The negative sign in front of $j^a$, $\ell_{\pm, 0}$ appears from the change in representation as differential operators on functions to variational action on fields \cite{Freedman:2012zz}.}
\be\ba{lll}\label{repJandL}
J^1 \= -j^1  +\frac{\i}{2} \delta_{M}({\lambda}_{2\tilde{1}} )\,,~&J^2 \= -j^2 
+\frac{\i}{2}\delta_{M}({\lambda}_{1\tilde{1}})\,,~&J^3 \= -j^3\,,
\\
L_+ \= -\ell_+  + \half \delta_M(  \i {\lambda}_{4\tilde{1}}+{\lambda}_{3\tilde{1}}  )\,,~&L_- \= -\ell_-  
+\half \delta_M( \i {\lambda}_{4\tilde{1}} -{\lambda}_{3\tilde{1}}  )\,,~&L_0 \= -\ell_0\,.
\ea\ee
Here,~$\delta_M(\hat\lambda_{a\tilde{b}})$ is the local Lorentz transformation in the~$\{ \cQ_a\,, \tilde{\cQ}_b\}$ 
algebra, as it appears in~\eqref{QQalgebra}, with field dependent parameters~$(\lambda_{a\tilde{b}})_{AB}\,.$\footnote{The $\delta_M(({\lambda}_{a \tilde{b}})_{AB})$ acts on a spinor~$\psi $ 
as~$\frac{1}{4}({\lambda}_{a \tilde{b}})_{AB} \gamma^{AB}\psi$ , and on a vector $V^A$ as $({\lambda}_{a \tilde{b}})^A_{~B} V^B$ .}  
On the background~(\ref{metric} - \ref{hyper_field_content}), their values are
\be\ba{lll}
(\lambda_{1 \tilde{1}})_{\hat{\theta}\hat{\phi}}\= 2 \frac{\sin\phi}{\sin\theta}  \,,~~&(\lambda_{2 \tilde{1}})_{\hat\theta \hat\phi}
\= 2 \frac{\cos\phi}{\sin\theta}  \,,\\
(\lambda_{3 \tilde{1}})_{\hat{t}\hat{\rho}}\= \frac{\cos(t +\psi)}{\cosh\rho}\,,~~~&(\lambda_{3 \tilde{1}})_{\hat{t}\hat{\psi}}
\=-\sin(t+\psi)\,,~~~&(\lambda_{3 \tilde{1}})_{\hat{\rho}\hat{\psi}}\=-\frac{\cos(t+\psi)}{\sinh\rho}\\
(\lambda_{4 \tilde{1}})_{\hat{t}\hat{\rho}}\= \frac{\sin(t +\psi)}{\cosh\rho}\,,~~~&(\lambda_{4 \tilde{1}})_{\hat{t}\hat{\psi}}
\=\cos(t+\psi)\,,~~~&(\lambda_{4 \tilde{1}})_{\hat{\rho}\hat{\psi}}\=-\frac{\sin(t+\psi)}{\sinh\rho}\,.
\ea
\ee

\bigskip

\section{Euclidean 4d supersymmetry and  AdS$_2 \times $ S$^2$} \label{4dSUSY}
In this appendix, we present the supersymmetry transformation of the fermions in Euclidean 4d 
conformal supergravity, following the convention of \cite{Jeon:2018kec}, and setting all fermions to zero.
The field content in Euclidean 4d superconformal gravity is given in Table~\ref{table:4dfieldcontent}. 
We also present the Euclidean AdS$_2 \times$ S$^2$ background and its Killing spinors.  
All fields appearing in this section refer to four-dimensional ones, so we omit the {4d} subscripts.

\medskip
\ndt

\begin{table}
\begin{center}
\renewcommand{\arraystretch}{1.5}
    \begin{tabular}{|c||c|c|c|}    
      \hline
      4d Weyl  & ${e_\mu}^{{a}}, \, \psi^i_a ,\,A^{D}_\mu , A^R_\mu\,,{\mathcal{V}_{\mu}}^i{}_j \,,\,{T}^{\pm}_{{a}{b}}\,,\, \cD,\,\chi^i_{4d}$ \\
      \hline
      4d Vector   & $X^{\mathcal{I}}\,,\overline{X}^{\mathcal{I}}\,,A^{\mathcal{I}}_\mu\,,\cY_{ij}^{\mathcal{I}},\, \lambda^{\cI\,i}$  \\
      \hline
       4d Hyper & ${\mathcal{A}_i}^{\alpha},\,\zeta^{\alpha}_{4d}$ \\
      \hline
      \hline
      4d SUSY parameters & $\epsilon^{i}_{4d},\,\eta^{i}_{4d}$\\
      \hline
      \end{tabular}
      \caption{Independent fields of the supersymmetric multiplets and $Q$, $S$-supersymmetry 
      parameters in four-dimensional $\mathcal{N}=2$ conformal supergravity. 
      } \label{table:4dfieldcontent}
   \end{center}
  \end{table}

\ndt
\textbf{Euclidean 4d supersymmetry transformations:}\\
The $Q$ and $S$-supersymmetry transformations of the fermionic fields are
\begin{equation}\label{QSKWeyl}
\begin{split}
\delta \psi_\mu^i & \= 2 D_\mu \varepsilon^i +\i\frac{1}{16} \gamma_{ab}(T^{ab+}+ T^{ab-})
\gamma_\mu  \varepsilon^i + \gamma_\mu \gamma_5 \eta^i\,,\\
 \delta \chi^i & \= \frac{\i}{24}\gamma_{ab}\slashed{D}(T^{ab+}+ T^{ab-})  \varepsilon^i 
 +\frac{1}{6}\widehat{R}(\cV)^i{}_{j \mu\nu}\gamma^{\mu\nu}\varepsilon^j - 
 \frac{1}{3} \widehat{R}(A^R)_{\mu\nu}\gamma^{\mu\nu}\gamma_5 \varepsilon^i \\
&~~~~~~~~~~~~~+ \cD\, \varepsilon^i +\i \frac{1}{24}(T^+_{ab}+ T^-_{ab})\gamma^{ab}\gamma_5\eta^i\,, \\
\delta\lambda_+^{ i}&\=- 2 \i \gamma^{a}D_{a}X{\varepsilon}_-^{i}-\half \cF_{ab}\gamma^{ab}
\varepsilon_+^{i}+\cY^{ij}\varepsilon_{jk}\varepsilon_+^{k}+2\i X\eta_+^{i}
\,,\\
\delta{\lambda}_-^{ i}&\= -2 \i \gamma^{a}D_{a}\overline{X}\varepsilon_+^{i}-\half \cF_{ab}
\gamma^{ab}{\varepsilon}_-^{i} +\cY^{ij}\varepsilon_{jk}{\varepsilon}_-^{k}-2\i\overline{X}{\eta}_-^{i}\,,\\
\delta \zeta^{\alpha} & \= \slashed{D}\cA_{i}{}^{\alpha}\ve^{i} 
- \cA_{i}{}^{\alpha}\gamma_5\eta^{i}\,,
\end{split}
\end{equation}
where:
\be\ba{l}
\cF_{\mu\nu}=F_{\mu\nu}-\left(\frac{1}{4} \overline{X}\,T^-_{\mu\nu}
+\frac{1}{4}  X\,{T}^+_{\mu\nu }
\right)
\,.
\label{fieldstrength}\ea\ee
The covariant derivatives are:
\begin{eqnarray}\label{Dve}
&&D_\mu \varepsilon^i=(\partial_{\mu}- \frac{1}{4}\omega_{\mu ab}\gamma^{ab} 
+\frac{1}{2}A^D_\mu  + \frac{1}{2}A^R_\mu \gamma_5)\varepsilon^i +\frac{1}{2}\cV_\mu{}^i{}_j \varepsilon^j\,,\\
&& D_{\mu}X \= (\partial_{\mu}-A^D_\mu +A^R_\mu)X\,, \\
&& D_{\mu}\overline{X} \= (\partial_{\mu}-A^D_\mu-A^R_{\mu})\overline{X}
 \,,\\
&& D_{\mu}\cA_{i}{}^{\alpha}=\left(\partial_{\mu}\cA_{i}{}^{\alpha}-b_\mu\right)\cA_{i}{}^{\alpha}+\half {{\cV_{\mu}}^j}_i \cA_{j}{}^{\alpha} \,,
\end{eqnarray}
and the curvatures are:
\begin{eqnarray} \label{curvatures}
&&\widehat{R}_{\mu\nu}(A^R)\= 2\partial_{[\mu}A^R_{\nu]}
\nonumber \,, \\
&&\widehat{R}_{\mu\nu}(\cV)^i{}_{j}\= 2\partial_{[\mu}\cV_{\nu]}{}^i{}_{j}+ \cV_{[\mu}{}^i{}_{k}\cV_{\nu]}{}^k{}_{j} \,.
\end{eqnarray}

\ndt
\textbf{Supersymmetric AdS$_2\times$S$^2$ background and Killing spinors:}  \label{KSonAdS2S2}

\ndt
Recall the fully supersymmetric, Euclidean AdS$_2\times$S$^2$ solution of the 4d theory considered in~\eqref{backgroundE}:
\beqa 
\label{backgroundEappendix}
&&ds^2= \R^2\left[d\eta^2+\sinh^2\eta\, d\chi^2+d\theta^2 +\sin^2\theta \,d\phi^2\right]\,,\\
&&F^{\cI}_{12}=-i\frac{ e^{\cI}}{\R^2}\,,~~~~~~F^{\cI}_{34}= \frac{p^{\cI}}{\R^2}\,,~~~ 
\leftrightarrow A^{\cI}= - \i e^{\cI} (\cosh \eta -1)d\chi  -p^{\cI} \cos\theta d\phi \\
&&{X}^{\cI} =\frac{\omega}{8}(e^{\cI} +ip^{\cI})\,,~~~~~\bar{X}^{\cI} =\frac{\bar{\omega}}{8}(e^{\cI} -ip^{\cI})\,,~~~{\cI}=0,1,\cdots, N_\text{v} 
\label{OnshellXappendix} \\
&&T^-_{12}=- \i\omega\,,~~~~~~~T^-_{34}=\i\omega \,,~~~~~~~{T}^+_{12}=-\i\bar{\omega}\,,~~~~~~~{T}^+_{34}=-\i \bar\omega\,.
\eeqa
Here, $\R$ is the radius of AdS$_2$ and S$^2$, and $\omega$,~$\overline{\omega}$ are two independent 
complex constants satisfying
\be \label{appendixomegarel}
\R^2=\frac{16}{\bar{\omega}\omega}\,.
\ee
As discussed in Section~\ref{connection}, we may pick the $SO(1,1)_R$ gauge~\eqref{omegamagn} such 
that~\eqref{appendixomegarel} implies the following parametrization:
\be
\omega(\alpha) \= \frac{4}{\ell}e^{\i \alpha}\,,\qquad \overline\omega(\alpha)  \= \ \frac{4}{\ell}e^{-\i \alpha }\,,\qquad 
\alpha \in \mathbb{R}\,.
\ee
Here, we choose $\alpha = \pi/2$ and derive the corresponding Killing spinors.

We express the AdS$_2 \times$ S$^2$ metric above in vielbein form:
\be
e^1= \ell \,d\eta\,,\quad e^2 =\ell \sinh\eta \,d\chi\,,\quad e^3 = \ell d \theta\,,\quad e^4 = \ell \sin\theta d \phi\,.
\ee
We also choose the following gamma matrix representation, where~$ \btau_a$ and~$\sigma_a$, $a=1,2,3$ are the Pauli matrices 
\be
\gamma_1= \btau_1\otimes \sigma_3\,,\quad \gamma_2= \btau_2\otimes \sigma_3\,,\quad \gamma_3= \mathbb{I}_2 
\otimes \sigma_1\,,\quad \gamma_4= \mathbb{I}_2 \otimes \sigma_2\,,\quad \gamma_5= \gamma_{1234}=- \btau_3 \otimes \sigma_3\,.
\ee
With this representation, the four-dimensional Killing spinor equation, given in~\eqref{QSKWeyl} as
\be
{\cal D}_\mu \ve = -\frac{\i}{32}(T^+_{ab}+ T^-_{ab})\gamma_{ab}\gamma_\mu \ve \=-\frac{1}{2\ell} 
(  \mathbb{I}_2 \times \sigma_3)\gamma_\mu \ve\,,
\ee
splits into the Killing spinor equations of AdS$_2$ and  S$^2$.
Indeed, decomposing the spinor $\ve= \ve_{\text{AdS}_2} \otimes \ve_{\text{S}^2}$, one obtains the AdS$_2$ part as
\be \label{AdS2spinors}
(\partial_\mu +\omega_\mu )\ve_{\text{AdS}_2}  \= -\frac{1}{2} \btau_\mu \, \ve_{\text{AdS}_2}\,,~~~\omega_\chi 
\= -\frac{\i}{2}\cosh\eta \,\btau_3\,,
\ee
and the S$^2$ as
\be \label{S2spinorsappendix}
(\partial_\mu +\omega_\mu )\ve_{\text{S}^2}  \= -\frac{1}{2}\sigma_3 {\sigma}_\mu \ve_{\text{S}^2}\,,~~~\omega_\phi 
\= -\frac{\i}{2}\cos\theta \,\sigma_3
\ee
The Killing spinors for AdS$_2$ and S$^2$ are given by
\be
\ve_{\text{AdS}_2}^+ = e^{\frac{\i}{2}\chi}\Biggl( \, \begin{matrix}- \cosh\frac{\eta}{2}\\ \sinh\frac{\eta}{2} \end{matrix}\, 
\Biggr)\,,~~~~ \ve_{\text{AdS}_2}^- = e^{-\frac{\i}{2}\chi}\Biggl( \, \begin{matrix} \sinh\frac{\eta}{2}\\ -\cosh\frac{\eta}{2} \end{matrix}\, \Biggr)\,,
\ee
and
\be
\ve_{\text{S}^2}^+ = e^{\frac{\i}{2}\phi}\Biggl( \, \begin{matrix} \cos\frac{\theta}{2}\\ \sin\frac{\theta}{2} 
\end{matrix}\, \Biggr)\,,~~~~\ve_{\text{S}^2}^- = e^{-\frac{\i}{2}\phi}\Biggl( \, \begin{matrix} \sin\frac{\theta}{2}\\ 
-\cos\frac{\theta}{2} \end{matrix}\, \Biggr)\,.
\ee
Taking the direct product of the spinors~\eqref{AdS2spinors} on AdS$_2$ with the 
spinors~\eqref{S2spinorsappendix} on S$^2$, we obtain the following complex basis of Killing spinors on AdS$_2\times$ S$^2$:
\be\ba{ll}\label{4complexKSin4d}
\dot\ve_+^{~+}= \sqrt{\frac{\ell}{2}}\,\varepsilon^+_{\text{AdS}_2}\otimes \varepsilon^{+}_{\text{S}^2} \,, 
\quad\quad\quad&\dot\varepsilon_+^{~-}= \sqrt{\frac{\ell}{2}}\,\varepsilon^+_{\text{AdS}_2}\otimes \varepsilon^{-}_{\text{S}^2} \,,\\
\dot\varepsilon_-^{~+}=\sqrt{\frac{\ell}{2}}\, \varepsilon^-_{\text{AdS}_2}\otimes \varepsilon^{+}_{\text{S}^2} \,, 
\quad\quad\quad&\dot\varepsilon_-^{~-}= \sqrt{\frac{\ell}{2}}\,\varepsilon^-_{\text{AdS}_2}\otimes \varepsilon^{-}_{\text{S}^2} \,.
\ea\ee
Note that, these spinors are identical to the Killing spinors on the Kaluza-Klein frame 
of~AdS$_3\times$S$^2$, given in \eqref{4KSKKframe}.
The spinors~\eqref{4complexKSin4d} organize themselves to form the following~8 real set 
of basis for Killing spinors on AdS$_2 \times$S$^2$,
\be\ba{ll}\label{8SMKSinKK}
\dot{\varepsilon}^{\,i}_{(1)}\=(-\i \dot\varepsilon_+^{~+}\,, \dot\varepsilon_-^{~-})\,, ~~~~~~&
\dot\varepsilon^{\,i}_{(2)}\=(\dot\varepsilon_+^{~+}\,,-\i\dot\varepsilon_-^{~-})\,,\\
\dot\varepsilon^{\,i}_{(3)}\=(-\dot\varepsilon_-^{~-}\,,-\i\dot\varepsilon_+^{~+})\,, &
\dot\varepsilon^{\,i}_{(4)}=(-\i\dot\varepsilon_-^{~-}\,,-\dot\varepsilon_+^{~+})\,,
\\
\dot{\tilde{\varepsilon}}^{\,i}_{(1)}\=(\dot\varepsilon_+^{~-}\,,\i\dot\varepsilon_-^{~+})\,,&
\dot{\tilde\varepsilon}^{\,i}_{(2)}\=(\i\dot\varepsilon_+^{~-}\,,\dot\varepsilon_-^{~+})\,,
\\
\dot{\tilde\varepsilon}^{\,i}_{(3)}\=(-\i\dot\varepsilon_-^{~+}\,,\dot\varepsilon_+^{~-})\,,&
\dot{\tilde\varepsilon}^{\,i}_{(4)}\=(\dot\varepsilon_-^{~+}\,,-\i\dot\varepsilon_+^{~-})\,,
\ea\ee
which is the same basis as for the 5d KK-frame \eqref{4SMKSinKK}. 
The spinors in~\eqref{8SMKSinKK} satisfy
\be\label{SM4dLorentzian}
(\ve^i)^\dagger \i \gamma_5 = \epsilon_{ij}(\ve^j)^T C\,.
\ee
which is indeed the reality condition, given in~\eqref{4drealityKS}, for~$\alpha=\pi/2$.

%

\providecommand{\href}[2]{#2}\begingroup\raggedright\endgroup

\end{document}